\documentclass[amsmath,notitlepage,amssymb,aps,showkeys,floatfix,prd,
a4paper,twocolumn,nofootinbib]{revtex4-2}

\usepackage{graphicx}
\usepackage{amsmath}
\usepackage{epstopdf}
\usepackage{amsfonts,amssymb}
\usepackage[utf8]{inputenc}
\usepackage{pstricks}
\usepackage{color}
\usepackage{multirow}
\usepackage{booktabs}
\usepackage{lipsum}

\usepackage[compat=1.0.0]{tikz-feynman}
\usepackage{simplewick}
\usepackage{color, colortbl}
\definecolor{Gray}{gray}{0.9}
\usepackage{float}
\usepackage{tikz-feynman}
\usepackage{feynmp}
\usepackage{forest}

\usepackage{verbatim}    
\usepackage{dcolumn}
\usepackage{bm}
\usepackage{epsfig}
\usepackage{braket}
\usepackage[normalem]{ulem} 
\usepackage{longtable} 

\newcommand{\be}{\begin{equation}}
\newcommand{\ee}{\end{equation}}
\newcommand{\ben}{\begin{eqnarray}}
\newcommand{\een}{\end{eqnarray}}

\def\MeV{\mbox{ MeV}}

\def\MeV{\mbox{ MeV}}

\newcommand{\pslash}{\not{\hbox{\kern-2.3pt $p$}}}
\newcommand{\pdslash}{\not{\hbox{\kern-2pt $\partial$}}}

\begin{document}

\title{Multiplicity of $Z_{cs}(3985)$  in heavy ion collisions }

\author{ L. M. Abreu}
%\email{luciano.abreu@ufba.br}
\affiliation{ Instituto de F\'isica, Universidade Federal da Bahia,
Campus Universit\'ario de Ondina, 40170-115, Bahia, Brazil\\
Instituto de F\'isica Corpuscular, Centro Mixto Universidad de Valencia-CSIC,
Institutos de Investigaci\'on de Paterna, Aptdo. 22085, 46071 Valencia, Spain }

\author{F. S. Navarra}
%\email{navarra@if.usp.br}
\affiliation{Instituto de F\'{\i}sica, Universidade de S\~{a}o Paulo, 
Rua do Mat\~ao, 1371 \\ CEP 05508-090,  S\~{a}o Paulo, SP, Brazil}

\author{M. Nielsen}
%\email{mnielsen@if.usp.br}
\affiliation{Instituto de F\'{\i}sica, Universidade de S\~{a}o Paulo, 
Rua do Mat\~ao, 1371 \\ CEP 05508-090,  S\~{a}o Paulo, SP, Brazil}

\author{H. P. L. Vieira}
%\email{hildeson.paulo@ufba.br}
\affiliation{ Instituto de F\'isica, Universidade Federal da Bahia,
Campus Universit\'ario de Ondina, 40170-115, Bahia, Brazil}

\begin{abstract}

  Using the coalescence model we compute the multiplicity of
  $Z_{cs} (3985)^-$ (treated as a compact tetraquark) at the end
  of the quark gluon plasma phase in heavy ion collisions. Then we
  study the time evolution of this state in the hot hadron gas phase.
  We calculate the thermal cross
  sections for the collisions of the $Z_{cs} (3985)^-$ with light mesons
  using effective Lagrangians and  form factors derived from 
  QCD sum rules for the vertices $Z_{cs}\bar{D}_{s }^* D $ and          
  $Z_{cs} \bar{D}_{s} D^{*}$. We solve the kinetic equation and find
  how the $Z_{cs} (3985)^-$ multiplicity is affected by the considered
  reactions during the expansion of the hadronic matter.
  A comparison with the statistical hadronization model predictions is 
  presented. Our results show that the tetraquark yield increases by a    
  factor of about $2-3$ from the hadronization to the kinetic freeze-out. 
  We also make predictions for the dependence of the $Z_{cs} (3985)^-$
  yield on the centrality, the center-of-mass energy and the charged
  hadron multiplicity measured at midrapidity
  $ \left[ d N_{ch} / d \eta (\eta < 0.5)\right]$.

\end{abstract}

\maketitle

%%%%%%%%%%%%%%%%%%%%%%%%%%%%%%%%%%%%%%%%%%%%%%%%%%%%%%%%%%%%%%%%%%%%%%%%%%%%%%%%
%%%%%%%%%%
\section{Introduction}
\label{Introduction} 
\begin{figure}[!ht]
%\hskip-8.0mm
  \includegraphics[{width=1.1\linewidth}]{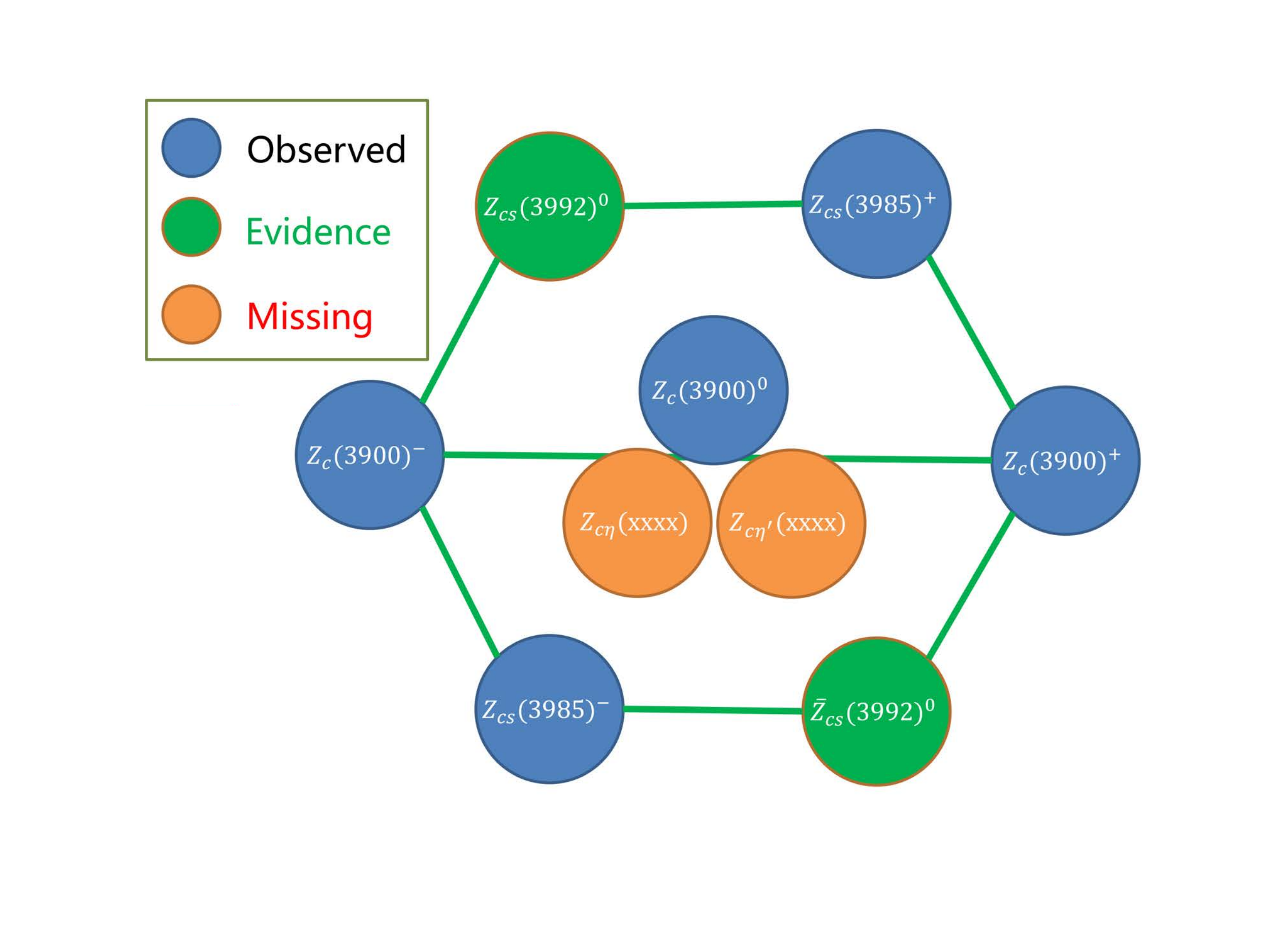}
  \vskip-3.0mm    
\caption{$J^P = 1^+$ $Z_c$ nonet formed by the product
  $ P \otimes J/\psi$. 
  Figure taken from \cite{yuan22}.}
\label{nonet}
\end{figure}

%%%%%%%%%%%%%%%%%%%%%%%%%%%%%%%%%%%%%%%%%%%%%%%%%%%%%%%%%%%%%%%%%%%%%%%%%%%%
%%%%%%%%%%
After 20 years of the $X(3872)$ observation, exotic charmonium
spectroscopy is still an exciting field with discoveries every year!
For recent reviews see \cite{yuan22,nora,zhu}.
As it happened in the case of ordinary light mesons, after the
discovery of several new particles a pattern emerged leading to the
successfull $SU(3)_F$ classification scheme. Over the past two decades
dozens of new hadronic states have been observed and now we need to
establish connections between all the new states. An example of
connection, relevant to this work, was suggested in \cite{yuan22},
where the members of the $Z_c$ (hidden charm states) family were
grouped into the $J^P = 1^+$ $Z_c$ nonet. This multiplet is obtained
combining the well known pseudoscalar $SU(3)_F$ nonet
$P = (\pi^-, \pi^0, \pi^+, K^-, K^+, K^0, \bar{K}^0, \eta, \eta')$
with a $J/\psi$, as shown Fig. \ref{nonet}. 
This organization of the states will eventually 
lead to an effective field theory of the $Z_c$ nonet interactions with the
light and heavy mesons. Such a theory is needed not only to explain the
structure and spectra of these states, but also to explain their production
mechanism in hadronic reactions.  

The recent observation of the $X(3872)$ in $Pb-Pb$ collisions by the CMS
Collaboration at the LHC \cite{cms22} opened a new era for the study of the
exotic charmonium states. In the beginning of these collisions,          
quark-gluon plasma (QGP) is formed. It then expands, cools down and
hadronizes into a hot hadron gas (HG), which lives for about $10$ fm and
finally freezes out giving origin to the observed particles. Hadrons are
formed at the beginning of the HG phase and hence interact with the (mostly
light) particles in the gas. From hadronization to freeze-out the
multiplicity of exotic charmonium changes because of its interactions with
light hadrons. Therefore, to understand the forthcoming data it is
crucial to have a reliable theory of these interactions. The $X(3872)$ has
been better studied and now there are several models which describe its 
interactions \cite{ChoLee1,wu-rapp21,nos14,XProd2}.
The interactions of the other states are much less known. In
\cite{Abreu:2022jmi} we developed an effective theory for the interactions
of the $Z_{cs} (3985)^-$ with light hadrons. In this work, we will apply it
to nucleus-nucleus collisions and will compute the $Z_{cs} (3985)^-$
multiplicity which may be measured in the future. 

At first sight, we might think that all the multiquark states produced at
the end of the quark-gluon plasma phase would be simply washed out during
their life in the hot hadron gas. This would be specially true if these
states were loosely bound meson molecules. However, our accumulated
experience has shown that, in most cases, these states are easily destroyed
but also easily produced. The result of this competition is
unpredictable. In some cases \cite{XProd2,nosups} the states are  
suppressed, in other cases their abundance remains approximately
constant \cite{noszb} or even grows slightly.
In \cite{Abreu:2022lfy}, we have shown that, in the case of the $T_{cc}^+$,
the time evolution of the abundance   depends on the internal structure
of the state: it grows if it is a compact tetraquark and drops if it is a
meson molecule.

Another naive guess would be the following: if the state has
a large hadronic decay width, it will be suppressed. Otherwise, it will
survive the hadron gas and be observed at the end of the collision. So far
this expectation has been confirmed by our calculations. Indeed, in
\cite{LeRoux:2021adw} we have shown that the $K^*$ is suppressed mainly  
because it decays ($K^* \to K \, \pi$ with a width of ~ $50$ MeV and
lifetime of ~ $4$ fm)  inside the hadron gas. On the other hand the
abundance of $D^*$ stays constant \cite{nosds} since
the decay $D^* \to D \, \pi$ (with a width of ~ $70$ keV) occurs only much
later, after the end of the hadron gas phase. The $Z_{cs} (3985)^-$ has
a decay width of ~ $12.8$ MeV, which corresponds to a lifetime of        
~ $15$ fm. Since this is longer than the expected lifetime of the hadron
gas, we will assume that the $Z_{cs} (3985)^-$  decays outside the hadronic
fireball and will neglect the effects of its width. 

In the next section, we present the $Z_{cs} (3985)^-$ 
(from now on simply $Z_{cs}$) interaction cross sections.
In Section III, we discuss the details of the rate equation which governs
the $Z_{cs}$  abundance. In Section IV we show our numerical results and
in the subsequent section we summarize our conclusions.

%%%%%%%%%%%%%%%%%%%%%%%%%%%%%%%%%%%%%%%%%%%%%%%%%%%%%%%%%%%%%%%%%%%%%%%%%%
%%%%%%%%
\section{The $Z_{cs}$ interactions}
\label{modelcrsec}
%%%%%%%%%%%%%%%%%%%%%%%%%%%%%%%%%%%%%%%%%%%%%%%%%%%%%%%%%%%%%%%%%%%%%%%%%%%%%%%%
%%%%%%%%

\begin{figure}[!ht]
\centering

\begin{tikzpicture}
\begin{feynman}
\vertex (a1) {$\bar{D}_s (p_1)$};
	\vertex[right=1.5cm of a1] (a2);
	\vertex[right=1.cm of a2] (a3) {$Z_{cs} (p_3)$};
	\vertex[right=1.4cm of a3] (a4) {$\bar{D}^{*}_{s} (p_1)$};
	\vertex[right=1.5cm of a4] (a5);
	\vertex[right=1.cm of a5] (a6) {$Z_{cs} (p_{3})$};
\vertex[below=1.5cm of a1] (c1) {$D (p_2)$};
\vertex[below=1.5cm of a2] (c2);
\vertex[below=1.5cm of a3] (c3) {$\pi (p_4)$};
\vertex[below=1.5cm of a4] (c4) {$D^{*} (p_2)$};
\vertex[below=1.5cm of a5] (c5);
\vertex[below=1.5cm of a6] (c6) {$\pi (p_4)$};
	\vertex[below=2cm of a2] (d2) {(a)};
	\vertex[below=2cm of a5] (d5) {(b)};
\diagram* {
  (a1) -- (a2), (a2) -- (a3), (c1) -- (c2), (c2) -- (c3), (c2) -- [fermion, edge
    label'= $D^{*}$] (a2), (a4) -- (a5), (a5) -- (a6), (c4) -- (c5), (c5) --
  (c6), (c5) -- [fermion, edge label'= $D$] (a5)}; 
\end{feynman}
\end{tikzpicture}

\begin{tikzpicture}
\begin{feynman}
\vertex (a1) {$\bar{D}_s (p_1)$};
	\vertex[right=1.5cm of a1] (a2);
	\vertex[right=1.cm of a2] (a3) {$Z_{cs} (p_3)$};
	\vertex[right=1.4cm of a3] (a4) {$D (p_1)$};
	\vertex[right=1.5cm of a4] (a5);
	\vertex[right=1.cm of a5] (a6) {$Z_{cs} (p_{3})$};
\vertex[below=1.5cm of a1] (c1) {$D^* (p_2)$};
\vertex[below=1.5cm of a2] (c2);
\vertex[below=1.5cm of a3] (c3) {$\pi (p_4)$};
\vertex[below=1.5cm of a4] (c4) {$\bar{D} (p_2)$};
\vertex[below=1.5cm of a5] (c5);
\vertex[below=1.5cm of a6] (c6) {$K (p_4)$};
	\vertex[below=2cm of a2] (d2) {(c)};
	\vertex[below=2cm of a5] (d5) {(d)};
\diagram* {
  (a1) -- (a2), (a2) -- (a3), (c1) -- (c2), (c2) -- (c3), (c2) -- [fermion, 
    edge label'= $D^{*}$] (a2), (a4) -- (a5), (a5) -- (a6), (c4) -- (c5),
  (c5) -- (c6), (c5) -- [fermion, edge label'= $D_{s}^{*}$] (a5)}; 
\end{feynman}
\end{tikzpicture}

\begin{tikzpicture}
\begin{feynman}
\vertex (a1) {$D^{*} (p_1)$};
	\vertex[right=1.5cm of a1] (a2);
	\vertex[right=1.cm of a2] (a3) {$Z_{cs} (p_3)$};
	\vertex[right=1.4cm of a3] (a4) {$D (p_1)$};
	\vertex[right=1.5cm of a4] (a5);
	\vertex[right=1.cm of a5] (a6) {$Z_{cs} (p_{3})$};
\vertex[below=1.5cm of a1] (c1) {$\bar{D}^{*} (p_2)$};
\vertex[below=1.5cm of a2] (c2);
\vertex[below=1.5cm of a3] (c3) {$K (p_4)$};
\vertex[below=1.5cm of a4] (c4) {$\bar{D}^* (p_2)$};
\vertex[below=1.5cm of a5] (c5);
\vertex[below=1.5cm of a6] (c6) {$K (p_4)$};
	\vertex[below=2cm of a2] (d2) {(e)};
	\vertex[below=2cm of a5] (d5) {(f)};
\diagram* {
  (a1) -- (a2), (a2) -- (a3), (c1) -- (c2), (c2) -- (c3), (c2) -- [fermion, 
    edge label'= $\bar{D}_{s}$] (a2), (a4) -- (a5), (a5) -- (a6), (c4) --
  (c5), (c5) -- (c6), (c5) -- [fermion, edge label'= $\bar{D}_{s}^{*}$]
  (a5)}; 
\end{feynman} 
\end{tikzpicture} 

\begin{tikzpicture}
\begin{feynman}
\vertex (a1) {$\bar{D}_s (p_1)$};
	\vertex[right=1.5cm of a1] (a2);
	\vertex[right=1.cm of a2] (a3) {$Z_{cs} (p_3)$};
	\vertex[right=1.4cm of a3] (a4) {$\bar{D}_s (p_1)$};
	\vertex[right=1.5cm of a4] (a5);
	\vertex[right=1.cm of a5] (a6) {$Z_{cs} (p_{3})$};
\vertex[below=1.5cm of a1] (c1) {$D_s (p_2)$};
\vertex[below=1.5cm of a2] (c2);
\vertex[below=1.5cm of a3] (c3) {$K (p_4)$};
\vertex[below=1.5cm of a4] (c4) {$D_s^{*} (p_2)$};
\vertex[below=1.5cm of a5] (c5);
\vertex[below=1.5cm of a6] (c6) {$K (p_4)$};
	\vertex[below=2cm of a2] (d2) {(g)};
	\vertex[below=2cm of a5] (d5) {(h)};
\diagram* {
  (a1) -- (a2), (a2) -- (a3), (c1) -- (c2), (c2) -- (c3), (c2) -- [fermion, 
    edge label'= $D^{*}$] (a2), (a4) -- (a5), (a5) -- (a6), (c4) -- (c5),
  (c5) -- (c6), (c5) -- [fermion, edge label'= $D^{*}$] (a5)};  
\end{feynman}
\end{tikzpicture}

\begin{tikzpicture}
\begin{feynman}
\vertex (a1) {$\bar{D}_s (p_1)$};
	\vertex[right=1.5cm of a1] (a2);
	\vertex[right=1.cm of a2] (a3) {$Z_{cs} (p_3)$};
	\vertex[right=1.4cm of a3] (a4) {$\bar{D}_s (p_1)$};
	\vertex[right=1.5cm of a4] (a5);
	\vertex[right=1.cm of a5] (a6) {$Z_{cs} (p_{3})$};
\vertex[below=1.5cm of a1] (c1) {$D (p_2)$};
\vertex[below=1.5cm of a2] (c2);
\vertex[below=1.5cm of a3] (c3) {$\eta (p_4)$};
\vertex[below=1.5cm of a4] (c4) {$D (p_2)$};
\vertex[below=1.5cm of a5] (c5);
\vertex[below=1.5cm of a6] (c6) {$\eta (p_4)$};
	\vertex[below=2cm of a2] (d2) {(i)};
	\vertex[below=2cm of a5] (d5) {(j)};
\diagram* {
  (a1) -- (a2), (a2) -- (a3), (c1) -- (c2), (c2) -- (c3), (c2) -- [fermion, 
    edge label'= $D^{*}$] (a2), (a4) -- (a5), (a5) -- (c6), (c4) -- (c5),
  (c5) -- (a6), (a5) -- [fermion, edge label'= $\bar{D}_{s}^{*}$] (c5)}; 
\end{feynman}
\end{tikzpicture}

\begin{tikzpicture}
\begin{feynman}
\vertex (a1) {$\bar{D}_{s}^{*} (p_1)$};
	\vertex[right=1.5cm of a1] (a2);
	\vertex[right=1.cm of a2] (a3) {$Z_{cs} (p_3)$};
	\vertex[right=1.4cm of a3] (a4) {$\bar{D}_{s}^{*} (p_1)$};
	\vertex[right=1.5cm of a4] (a5);
	\vertex[right=1.cm of a5] (a6) {$Z_{cs} (p_{3})$};
\vertex[below=1.5cm of a1] (c1) {$D^{*} (p_2)$};
\vertex[below=1.5cm of a2] (c2);
\vertex[below=1.5cm of a3] (c3) {$\eta (p_4)$};
\vertex[below=1.5cm of a4] (c4) {$D^{*} (p_2)$};
\vertex[below=1.5cm of a5] (c5);
\vertex[below=1.5cm of a6] (c6) {$\eta (p_4)$};
	\vertex[below=2cm of a2] (d2) {(k)};
	\vertex[below=2cm of a5] (d5) {(l)};
\diagram* {
  (a1) -- (a2), (a2) -- (a3), (c1) -- (c2), (c2) -- (c3), (c2) -- [fermion, 
    edge label'= $D$] (a2), (a4) -- (a5), (a5) -- (c6), (c4) -- (c5),
  (c5) -- (a6), (a5) -- [fermion, edge label'= $\bar{D}_{s}$] (c5)}; 
\end{feynman}
\end{tikzpicture}

\begin{tikzpicture}
\begin{feynman}
\vertex (a1) {$\bar{D}_{s}^{*} (p_1)$};
	\vertex[right=1.5cm of a1] (a2);
	\vertex[right=1.cm of a2] (a3) {$Z_{cs} (p_3)$};
\vertex[below=1.5cm of a1] (c1) {$D (p_2)$};
\vertex[below=1.5cm of a2] (c2);
\vertex[below=1.5cm of a3] (c3) {$\eta (p_4)$};
	\vertex[below=2cm of a2] (d2) {(m)};
\diagram* {
  (a1) -- (a2), (a2) -- (c3), (c1) -- (c2), (c2) -- (a3), (a2) -- [fermion,
    edge label'= $\bar{D}_{s}^{*}$] (c2)}; 
\end{feynman}
\end{tikzpicture}

\caption{Diagrams contributing to the following processes (without
  specification 
  of the charges of the particles): $ \bar{D}_{s}^{(*)} D^{(*)}
  \rightarrow Z_{cs} \pi $  [(a)-(c)], $ D^{(*)} \bar{D}^{(*)} , \bar{D}_{s}
  D_s^{(*)}  \rightarrow Z_{cs} K $ [(d)-(h)] and $ \bar{D}_{s}^{(*)}
  D^{(*)}
  \rightarrow Z_{cs}\eta $  [(i)-(m)]. They are reproduced from
  Ref.~\cite{Abreu:2022jmi}. }
\label{DIAG1}
\end{figure}
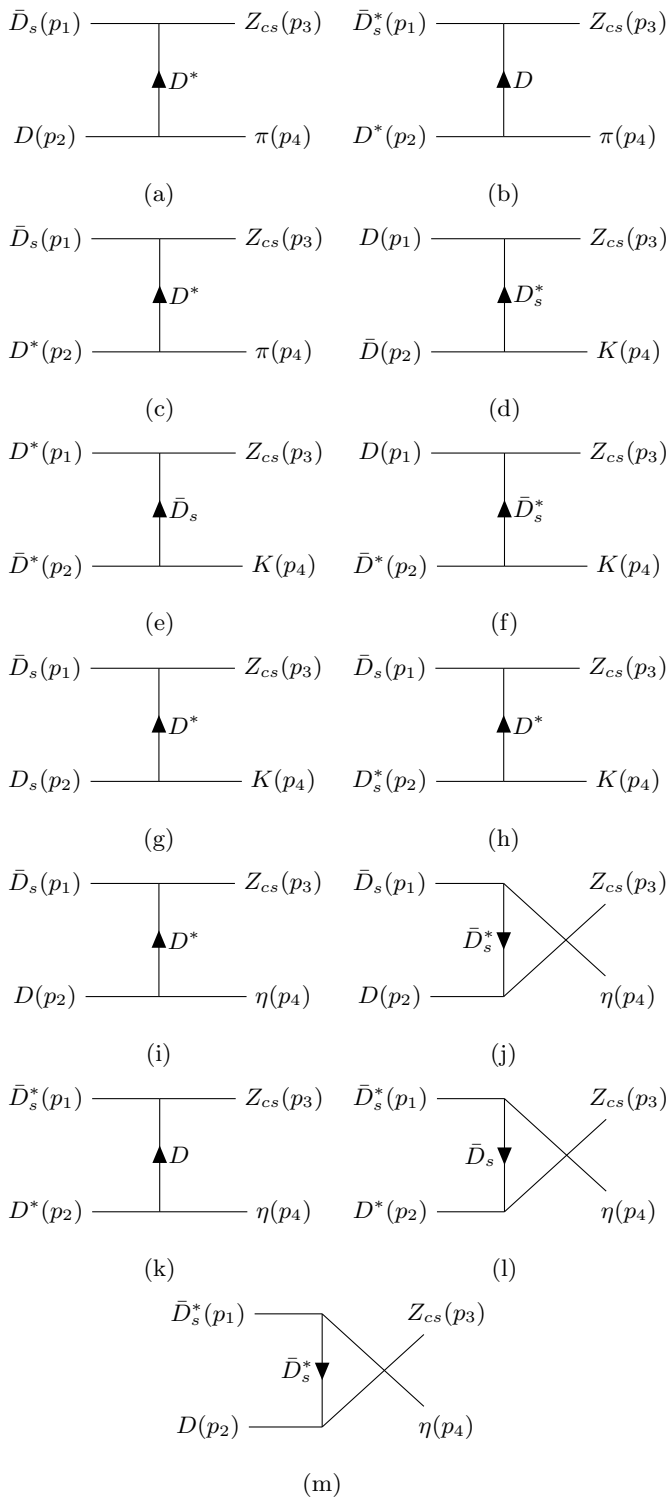

In this work we will focus on the multiplicity of the $Z_{cs}$ produced 
in heavy-ion collisions. The $Z_{cs}$ interactions with the lightest 
pseudoscalar mesons have been addressed in our previous work
~\cite{Abreu:2022jmi}, where the thermal  cross sections for the reactions 
$  \bar{D}_{s}^{(*)} D^{(*)}  \rightarrow Z_{cs} \pi $, $  D^{(*)} 
\bar{D}^{(*)} , \bar{D}_{s} D_s^{(*)} \rightarrow Z_{cs} K $  and    
$ \bar{D}_{s}^{(*)} D^{(*)} \rightarrow Z_{cs}\eta $, as well as for the
inverse processes, have been calculated. In Fig.~\ref{DIAG1} we show the    
lowest-order Born diagrams of the relevant processes. To calculate these
cross  
sections, we have used an effective theory approach, with the couplings 
involving $\pi$, $K^{(*)}$, $D^{(*)}$ and $D_s^{(*)}$ mesons based on     
pseudoscalar-pseudoscalar-vector and vector-vector-pseudoscalar vertices.   
The couplings involving the $Z_{cs}$ have been introduced assuming
that this is a $S$-wave state engendered by the superposition of
$  D_{s}^{* -} D^{0} $ and $  D_{s}^{-} D^{* 0} $  configurations, with   
quantum numbers $I (J^P) = \frac{1}{2} (1^+) $. This can be represented by
the effective Lagrangian~\cite{Wu:2021ezz},   
\begin{eqnarray}\label{Lagr3}
  \mathcal{L}_{Z_{cs}} = \frac{ g_{Z_{cs}} }{ \sqrt{2} }
  Z_{cs}^{\dagger \mu}
  ( \bar{D}_{s \mu}^{*} D + \bar{D}_{s} D_{\mu}^{*} ),
\end{eqnarray}
where  $Z_{cs}$ denotes the field associated to the $Z_{cs} (3985)^-$ state. 
Also, the $\bar{D}_{s \mu}^{*} D $ and
$\bar{D}_{s} D_{\mu}^{*}$ mean the $D_{s}^{* -} D^{0}$ and
$D_{s}^{-} D^{* 0}$ components, respectively.  

Two aspects of the formalism developed   in
Ref.~\cite{Abreu:2022jmi} deserve special comments. The first one concerns  
the effective coupling constant $ g_{Z_{cs}} $, whose value has been taken 
from Ref.~\cite{Wu:2021ezz}. In that paper  $g_{Z_{cs}} $ has been 
estimated to be $ 6.0 - 6.7$ GeV
%in order to properly describe the 
%$Z_{cs}$ total width for the decay channels                      
%$Z_{cs} \rightarrow D_{s}^{*} \bar{D} + h.c., J/\psi K, \eta K $,
assuming that the $Z_{cs}$ is a $S$-wave molecule of               
$ ( \bar{D}_{s \mu}^{*} D + \bar{D}_{s} D_{\mu}^{*} ) $. The second 
aspect refers to the 
empirical monopole-like form factor that has been introduced  in    
Ref.~\cite{Abreu:2022jmi} to account for the composite nature of hadrons
and their finite extension and also to avoid the artificial growth of the
amplitudes with the energy.

In this work we will assume that the $Z_{cs}$ is a compact tetraquark. The
interaction Lagrangian (\ref{Lagr3}) is the same used to study molecular   
states but the coupling constant is different and can be determined with
QCD sum rules (QCDSR). 
%The reason resides on the fact that the QCDSR allows one to determine 
%analytically (and approximately) the form factor from the three-point  
%correlation function, which is evaluated in two ways. Firstly from the 
%quark currents, written in terms of the flavor and color content with  
%the correct quantum numbers, i.e. the Operator Product Expansion (OPE)
%description. The other is to treat the correlation function in terms of
%matrix elements of hadronic states, estimated with effective Lagrangians,
%i.e. the Phenomenological description. Then, the matching between these
%two evaluations engenders a function, the so-called form factor.
For a detailed discussion on this issue, we refer the reader to
Ref.~\cite{Abreu:2021jwm}.
The advantage of using the QCDSR method is that it is more firmly rooted
in QCD and also  it is more appropriate to the study of multiquark
systems in a compact configuration, as it might be the case of the
$Z_{cs}$ state. In Ref.~\cite{Dias:2013qga} a detailed
analysis of the three-point correlation function of the vertices 
$Z_{cs}\bar{D}_{s }^* D $ and $Z_{cs} \bar{D}_{s} D^{*}$ was performed  
and the relevant form factor was numerically calculated and  parametrized
by the exponential function:
\be
F_{Z_{cs}}(Q^2) = g_1 \, e^{-g_2 Q^2},
\label{exp}
\ee
where $Q^2 = -q^2$ is the Euclidean four-momentum of the off-shell    
particle (the exchanged $ D_s $ meson), $ g_1 =0.94~\mbox{GeV}$    
and $ g_2 =0.08~\mbox{GeV}^{-2}$.
The coupling constant was obtained from the value of the form factor
at the meson pole:
%%%%%%%%%%%%%%%%%%%%%
\be
g_{Z_{cs}}=F_{Z_{cs}}(Q^2 = -m^2_{D_s}) = (1.4 \pm 0.4)~~\mbox{GeV}.
\label{coupdd}
\ee
%%%%%%%%%%%%%%%%%%%%%
\begin{widetext}

\begin{figure}[!ht]
    \centering
\includegraphics[{width=0.32\linewidth}]{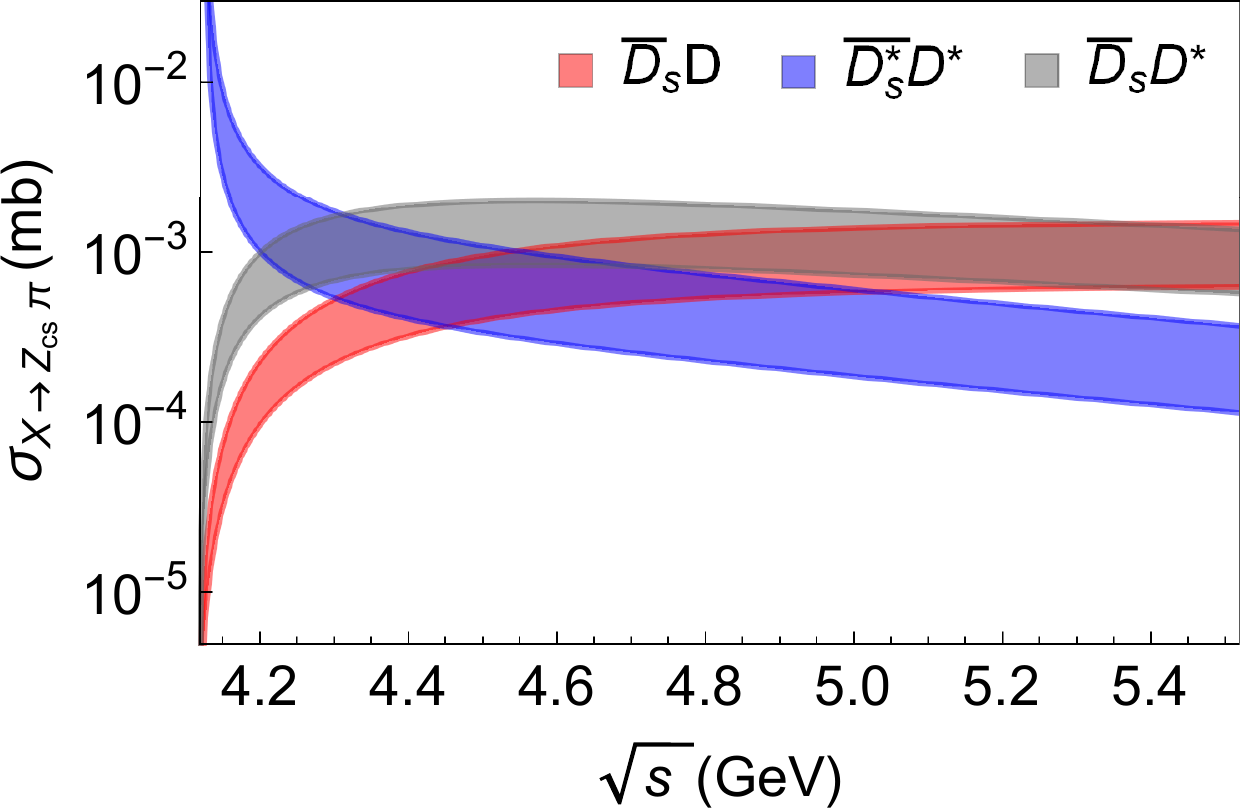}
\includegraphics[{width=0.32\linewidth}]{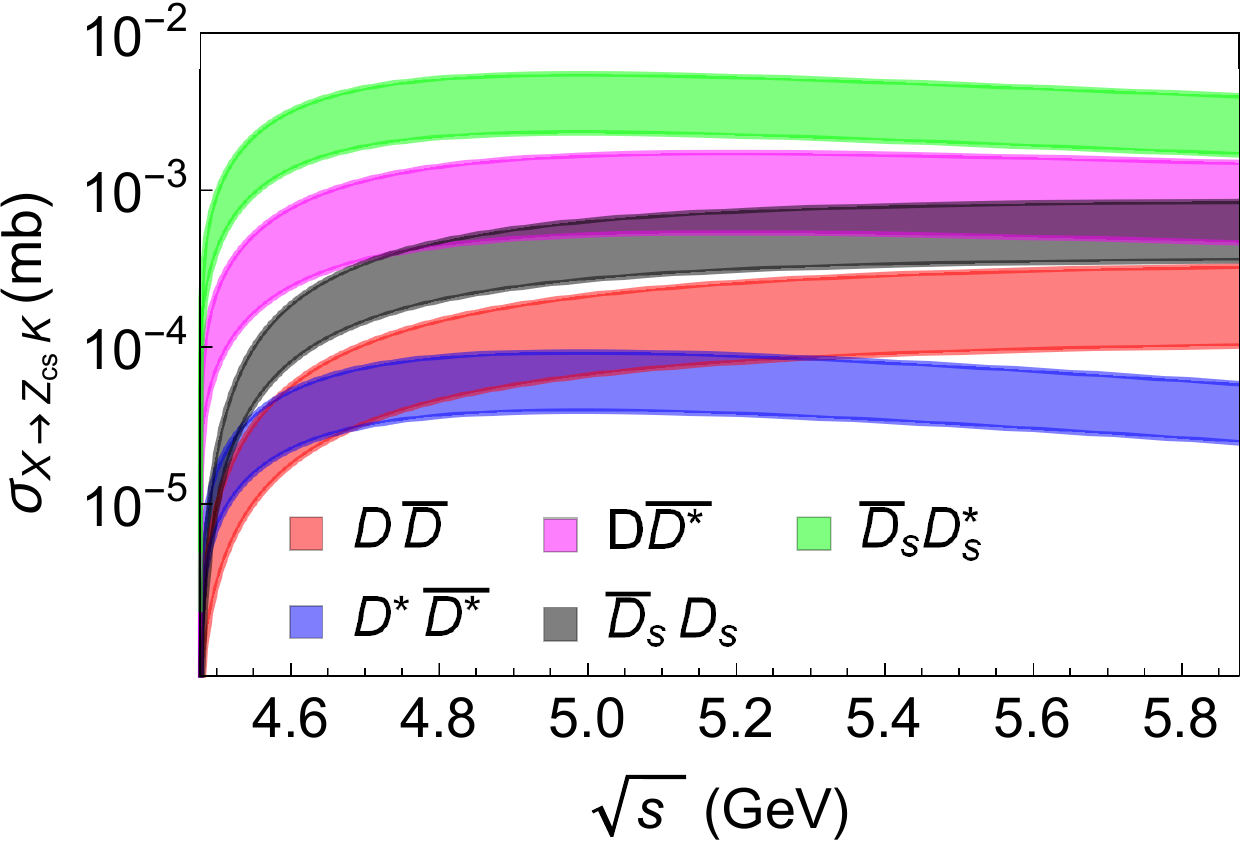}
\includegraphics[{width=0.32\linewidth}]{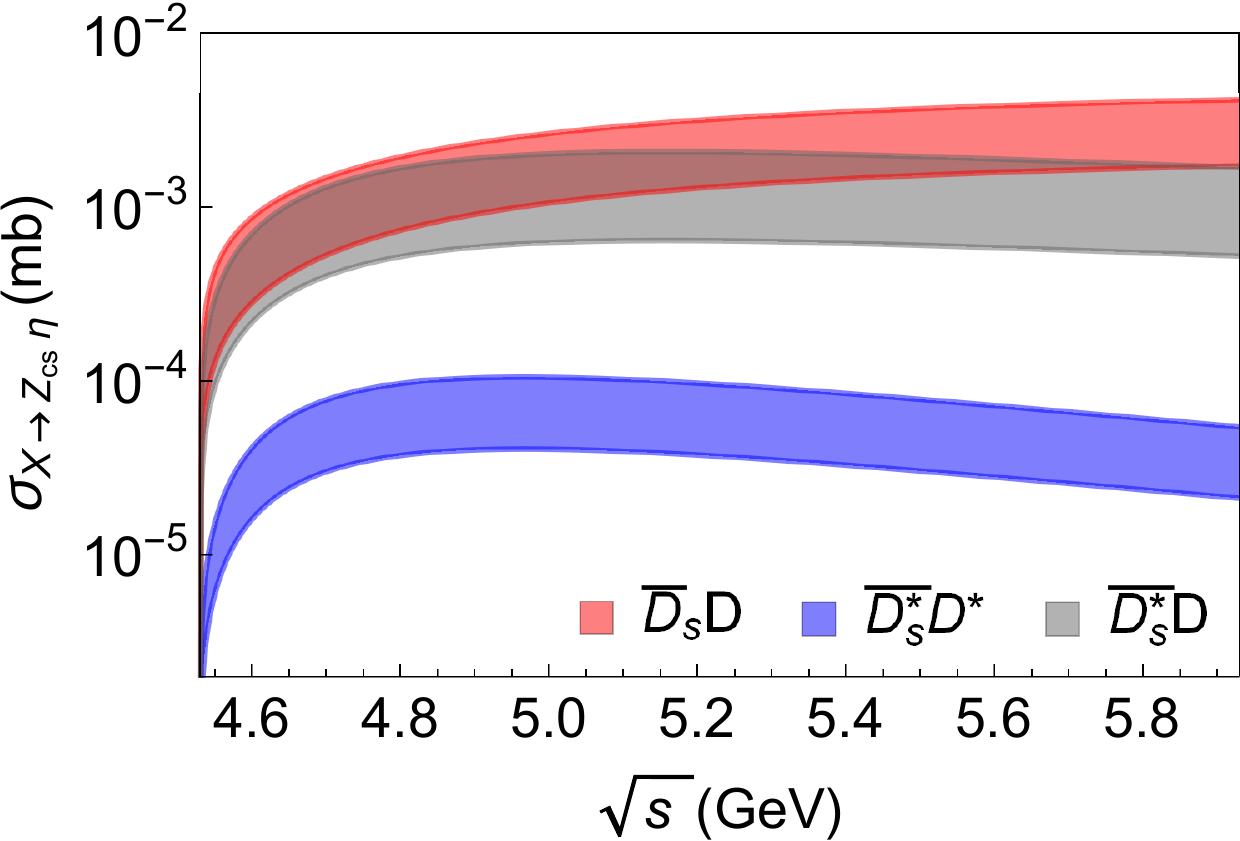}
\caption{Cross sections for the production processes  $ Z_{cs}^{-}\pi $
  (left),
  $Z_{cs}^{-}K$(center) and $Z_{cs}^{-}\eta$(right), as  functions of  
  center-of-mass energy $\sqrt{s}$. The bands in the plots denote the
  uncertainties considered in Eq.~(\ref{coupdd}). }
    \label{Fig:CrSec-Prod}
\end{figure}
%aqui
\end{widetext}
We  use the form factor~(\ref{exp}) and the coupling        
constant~(\ref{coupdd}) for the vertices  $Z_{cs}\bar{D}_{s }^* D $      
and $Z_{cs} \bar{D}_{s} D^{*}$  to calculate the cross sections 
of the processes displayed in Fig.~\ref{DIAG1}. The explicit expressions 
for the amplitudes and cross sections can be found in our previous  
work~\cite{Abreu:2022jmi} and the resulting cross sections are shown in
Fig.~\ref{Fig:CrSec-Prod}.   The cross sections obtained in
~\cite{Abreu:2022jmi} are different from those shown in
Fig.~\ref{Fig:CrSec-Prod}. The former were obtained with monopole-like
form factors and $g_{Z_{cs}}= 6.0 - 6.7$ GeV. The latter were obtained with  
QCDSR form factors and $g_{Z_{cs}}= 1.4$ GeV. The new cross sections are one
order of magnitude smaller (than those found in
~\cite{Abreu:2022jmi}), they fall more slowly as the center-of-mass
energy increases and they have large theoretical errors. These features
can be attributed to (\ref{exp}), (\ref{coupdd}) and the associated errors,
discussed in \cite{Dias:2013qga}. The QCDSR calculation can be made more
precise if one includes more terms in the operator product expansion and
if one has better experimental information on the $Z_{cs}$ decays. 

In the hot hadronic medium formed in heavy-ion the temperature
drives the collision energy. Therefore we need to evaluate the
thermally averaged cross sections (or simply thermal cross sections),
defined as convolutions of the vacuum cross sections with the thermal
momentum distributions of the colliding particles. For processes with a
two-particle initial state going into two final particles $ab \to cd$,
it is given by~\cite{Koch,ChoLee1,XProd2}: 
\begin{eqnarray}
  \langle \sigma_{a b \rightarrow c d } \,
  v_{a b}\rangle &  = & \frac{ \int 
    d^{3} \mathbf{p}_a  d^{3} \mathbf{p}_b \, f_a(\mathbf{p}_a) \,
    f_b(\mathbf{p}_b) \, 
    \sigma_{a b \rightarrow c d } \,\,v_{a b} }{ \int  d^{3} \mathbf{p}_a
    d^{3}
    \mathbf{p}_b \, f_a(\mathbf{p}_a) \,  f_b(\mathbf{p}_b) }
\nonumber \\
& = & \frac{1}{4 \, \alpha_a ^2 \, K_{2}(\alpha_a) \, \alpha_b ^2 \,
  K_{2}(\alpha_b) }
\int _{z_0} ^{\infty } dz \,  K_{1}(z) \,\,
\nonumber \\
& & \times \sigma (s=z^2 T^2) \left[ z^2 - (\alpha_a + \alpha_b)^2 \right]
\nonumber \\
& &\left[ z^2 - (\alpha_a - \alpha_b)^2 \right],
\label{ThermalCrossSection}
\end{eqnarray}
where $v_{ab}$ denotes the relative velocity of the two initial  interacting 
particles $a$ and $b$; $\sigma_{ab \to cd}$  represents the cross sections 
for the different reactions shown in Fig.~\ref{DIAG1}; the function
$f_i(\mathbf{p}_i)$ is the Bose-Einstein   
distribution of particles of species $i$, which depends on the temperature
$T$; $\beta _i = m_i / T$, $z_0 = max(\beta_a + \beta_b,\beta_c 
+ \beta_d)$; and  $K_1$ and $K_2$ the modified Bessel functions.

\begin{widetext}

  \begin{figure}[!ht]
\vskip3.0mm    
    \centering
\includegraphics[{width=0.32\linewidth}]{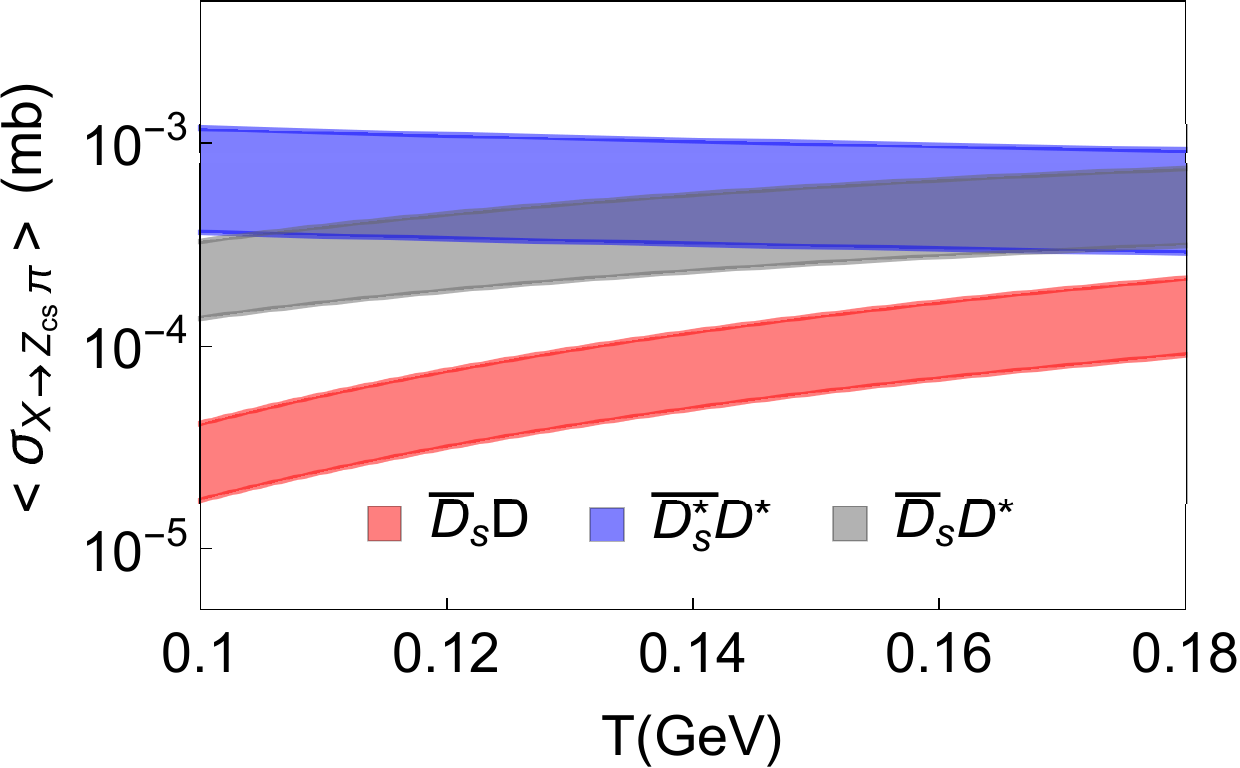}
\includegraphics[{width=0.32\linewidth}]{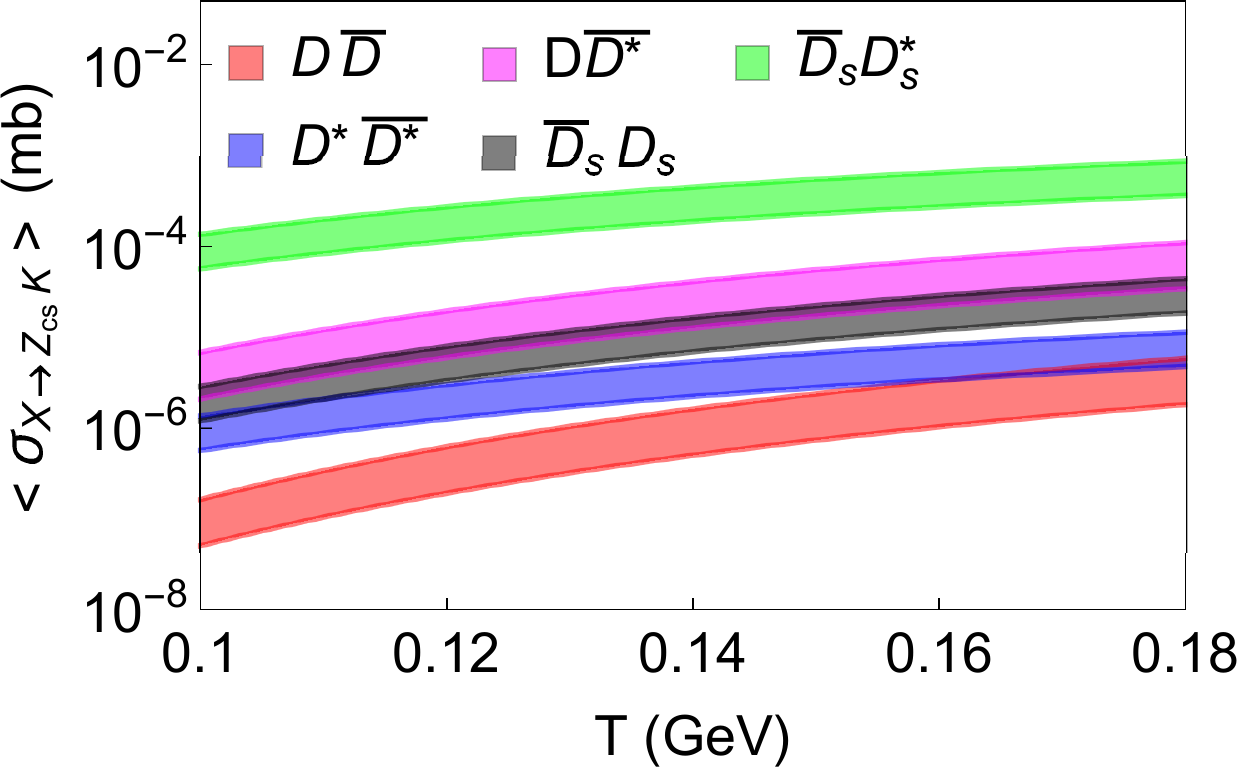}
\includegraphics[{width=0.32\linewidth}]{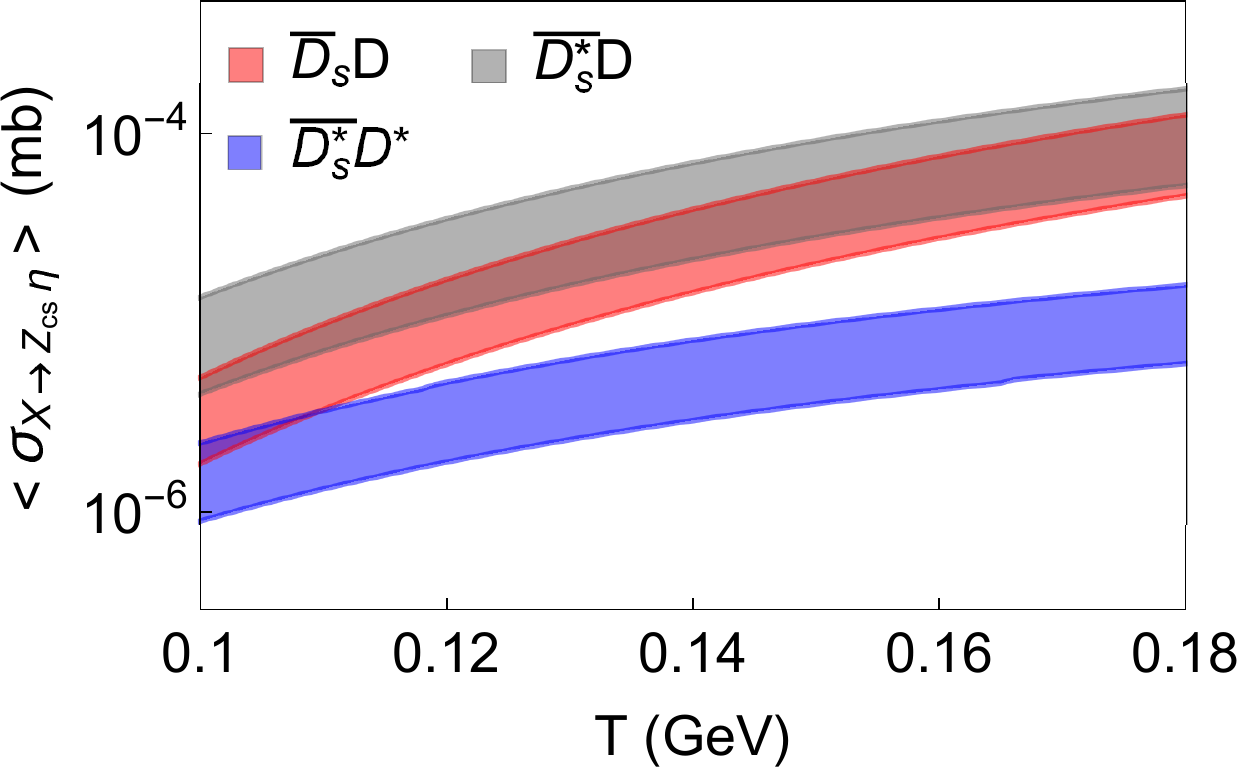}\\
\includegraphics[{width=0.32\linewidth}]{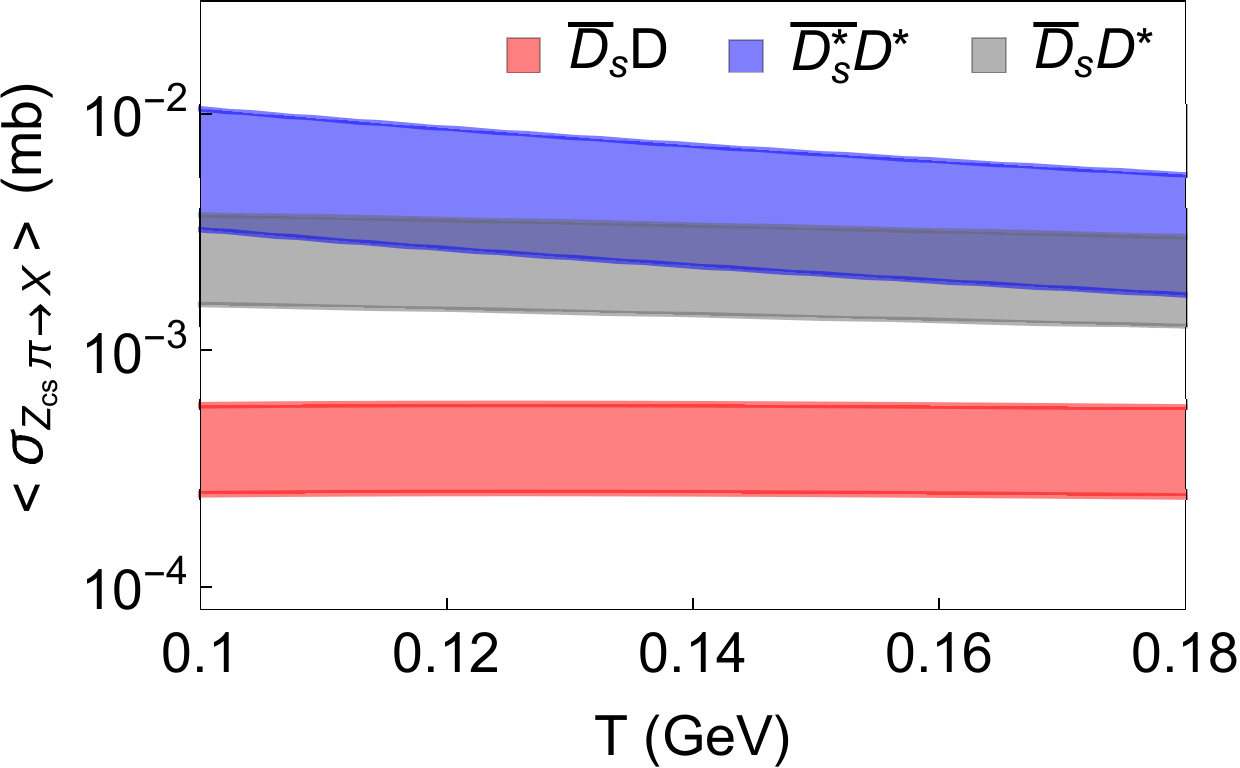}
\includegraphics[{width=0.32\linewidth}]{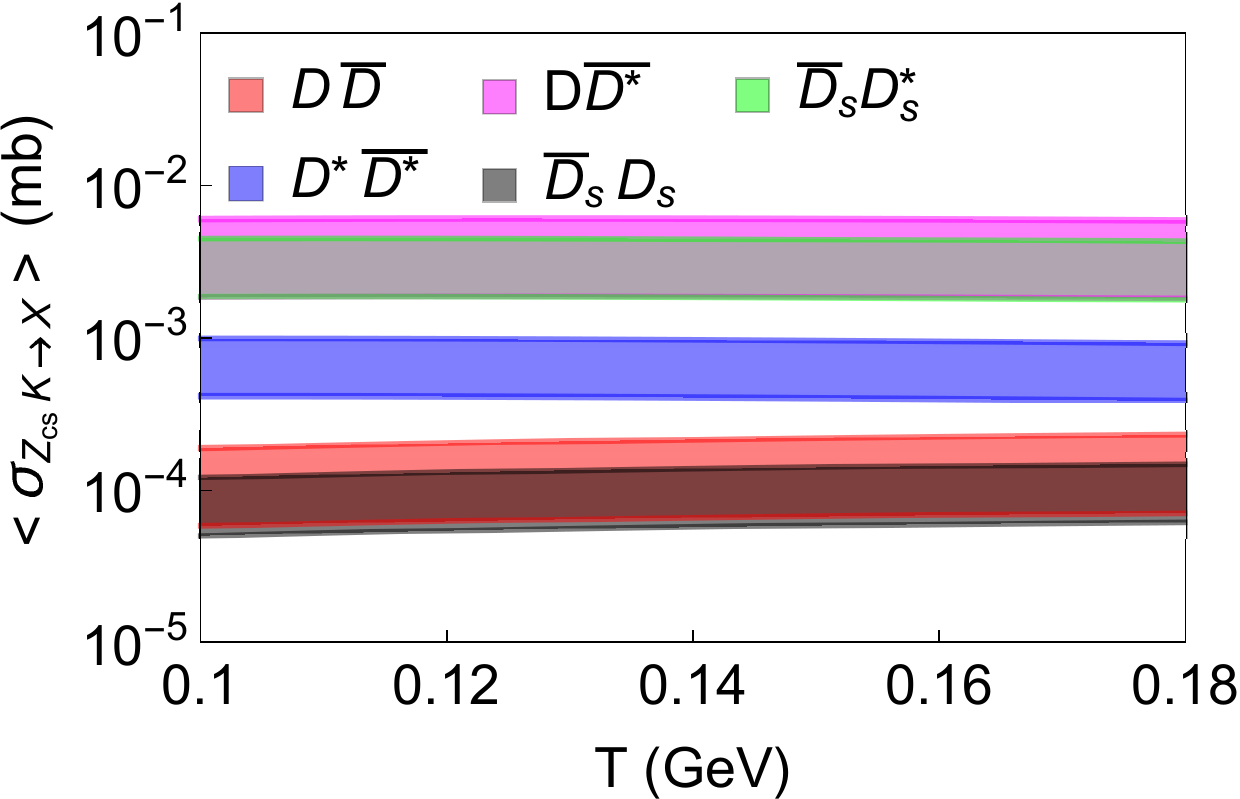}
\includegraphics[{width=0.32\linewidth}]{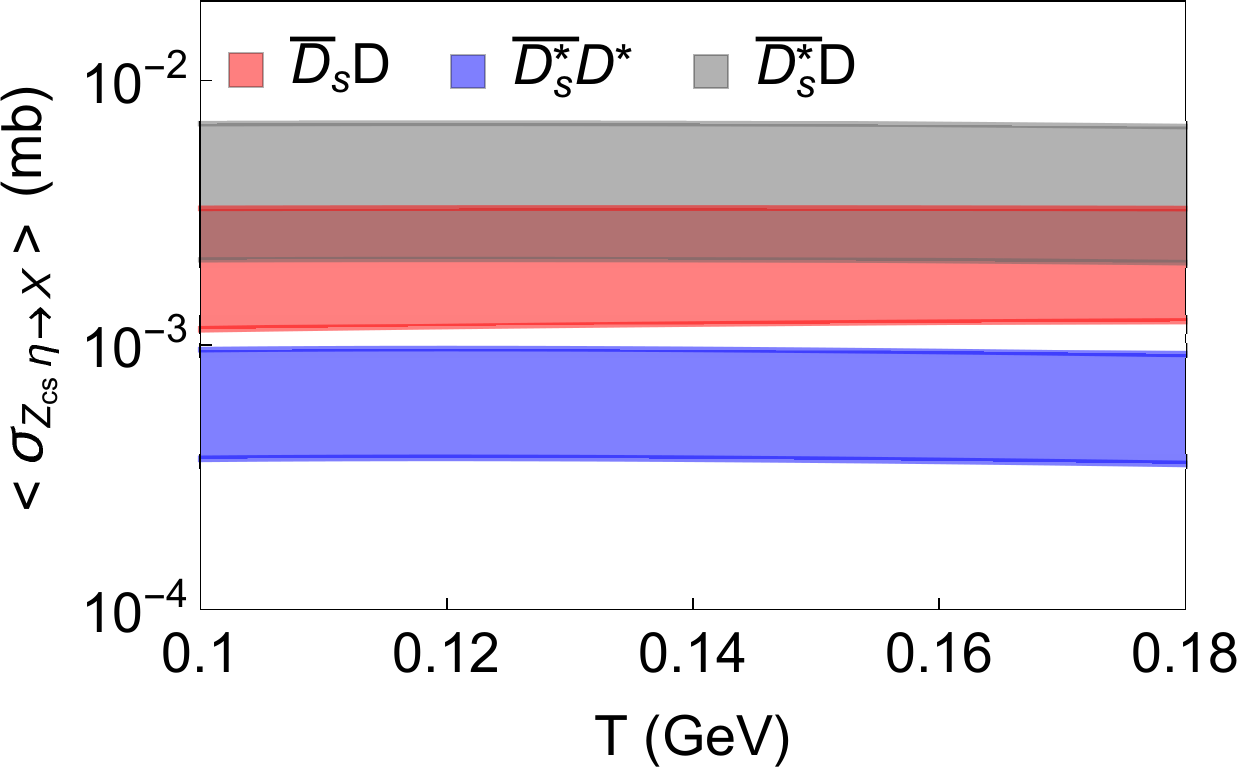}
  \caption{Top: thermal cross sections for the production processes         
$Z_{cs}^{-}\pi$(left), $Z_{cs}^{-}K$(center) and $Z_{cs}^{-}\eta$(right), 
as a function of temperature $T$. Bottom: the same as in the top panel
but for the corresponding suppression processes. }
    \label{Fig:AvCrSec}
\end{figure}

\end{widetext}

In Fig.~\ref{Fig:AvCrSec} we show the   
thermal cross sections for $Z_{cs}$ production and absorption plotted as    
functions of the temperature. The results for  $Z_{cs}$ absorption have
been evaluated by using the detailed balance relation.
In general, the results reveal the same qualitative behavior of those 
discussed in~\cite{Abreu:2022jmi}: the thermal cross sections for the 
$Z_{cs}$ absorption do not change much in  this range of temperature,
staying almost constant. On the other hand, in the case of $Z_{cs}$
production, most of the cross sections grow significantly with the
temperature. 

The qualitative differences between the results shown in
Fig.~\ref{Fig:AvCrSec} (obtained with QCDSR) and the corresponding ones
reported in~\cite{Abreu:2022jmi} (obtained with an empirical form factor)
remain the same: the QCDSR based results are smaller than the empirical
ones by one order of magnitude. In addition, the larger bands in the plots
are due to the greater relative uncertainties. 

Most importantly, as in~\cite{Abreu:2022jmi} the thermal cross sections 
for  $Z_{cs}$ absorption are greater than those for production at least      
by one order of magnitude, depending on the temperature. This result     
might have a considerable impact on the observed  $Z_{cs}$ multiplicity  
in heavy ion collisions: the initial yield at the end of the quark-gluon  
plasma phase might suffer significant changes during the hadron gas phase 
because of the interactions. This is what will be investigated in the
next sections. 

%%%%%%%%%%%%%%%%%%%%%%%%%%%%%%%%%%%%%%%%%%%%%%%%%%%%%%%%%%%%%%%%%%
\section{The  $Z_{cs}$ multiplicity }
\label{model}
%%%%%%%%%%%%%%%%%%%%%%%%%%%%%%%%%%%%%%%%%%%%%%%%%%%%%%%%%%%%%%%%%%
\subsection{The time evolution}
\label{time}
%%%%%%%%%%%%%%%%%%%%%%%%%%%%%%%%%%%%%%%%%%%%%%%%%%%%%%%%%%%%%%%%%%

We are interested in the effect of the
$D_{(s)}^{(*)} D_{(s)}^{(*)} \leftrightarrow Z_{cs}\pi$, $Z_{cs}K$ and      
$Z_{cs}\eta$  interactions on the abundance of $Z_{cs}$ during  
the hadron gas phase of heavy ion collisions. To this end, we model the    
time evolution of $Z_{cs}$ multiplicity in the same way as 
in Ref.~\cite{Abreu:2022lfy}, through the momentum-integrated rate
equation ~\cite{ChoLee1,XProd2,Koch}: 
\begin{eqnarray} 
\frac{ d N_{Z_{cs}} (\tau)}{d \tau} & = & \sum_{\substack{c, c^{\prime} = D,
      D^* \\ \varphi = \pi, K, \eta }} 
\left[ \langle \sigma_{c c^{\prime} \rightarrow Z_{cs} \varphi } 
v_{c c^{\prime}} \rangle n_{c} (\tau) N_{c^{\prime}}(\tau)
 \right. \nonumber \\ & &  \left. 
 - \langle \sigma_{ \varphi Z_{cs} \rightarrow c c^{\prime} }
 v_{ Z_{cs} \varphi } 
\rangle n_{\varphi} (\tau) N_{Z_{cs}}(\tau) 
%\right. \nonumber \\  & &   \left. 
%{\color{red} - \Gamma_{Z_b} N_{Z_b} (\tau) } 
\right], 
\label{rateeq}
\end{eqnarray}
where $N_{Z_{cs}} (\tau)$, $N_{c^{\prime}}(\tau)$,  $ n_{c} (\tau)$ and  
$n_{\varphi} (\tau)$ are the abundances of $Z_{cs}$ and of              
charmed (strange) mesons of type $c^{(\prime )}$, and the densities of 
charmed (strange) mesons of type $c ^{(\prime )}$ and of light mesons
$\varphi $  at proper time $\tau$, respectively. 

In order to solve this equation we need to define the initial condition
and we need to know how the quantities on the right side of the equation
depend on time. This will be discussed in the next sections.

%%%%%%%%%%%%%%%%%%%%%%%%%%%%%%%%%%%%%%%%%%%%%%%%%%%%%%%%%%%%%%%%%%%%
%\begin{itemize}
%\item \textit{Medium in equilibrium. }
\subsubsection{Medium in equilibrium}
%%%%%%%%%%%%%%%%%%%%%%%%%%%%%%%%%%%%%%%%%%%%%%%%%%%%%%%%%%%%%%%%%%%%

The light and heavy mesons in the medium are assumed to be in thermal
equilibrium and hence  $ n_{c} (\tau)$, $n_{c^{\prime}}(\tau)$ and        
$n_{\varphi} (\tau)$ can be written in the Maxwell-Boltzmann approximation
as~\cite{ChoLee1,XProd2,Koch}
\ben
n_{i} (\tau) &  \approx & \frac{1}{2 \pi^2}\ \gamma_{i} \ g_{i}
\ m_{i}^2  \ T(\tau) \ K_{2}\left(\frac{m_{i} }{T(\tau)}\right), 
\label{densities}
\een
where $\gamma _i$, $g_i$,  $m_i$ are the fugacity, the degeneracy  
factor and the mass  of the particle of type $i$, respectively. The  
product of the density $n_i(\tau)$ by the volume $V(\tau)$ gives the
multiplicity $N_i (\tau)$.

%\item \textit{Boost invariant Bjorken picture.}
%%%%%%%%%%%%%%%%%%%%%%%%%%%%%%%%%%%%%%%%%%%%%%%%%%%%%%%%%%%%%%%%%%%%%%
\subsubsection{Hydrodynamical expansion }
%%%%%%%%%%%%%%%%%%%%%%%%%%%%%%%%%%%%%%%%%%%%%%%%%%%%%%%%%%%%%%%%%%%%%%

The relevant quantities depend on time through the temperature
$T(\tau)$ and  volume $V(\tau)$, which are parametrized  in order 
to simulate the boost invariant  Bjorken expansion of the hadron gas 
\cite{ChoLee1,XProd2,Koch} : 
\ben
V(\tau) & = & \pi \left[ R_C + v_C \left(\tau - \tau_C\right) + 
\frac{a_C}{2} \left(\tau - \tau_C\right)^2 \right]^2 \tau c , \nonumber \\
T(\tau) & = & T_C - \left( T_H - T_F \right) \left( \frac{\tau - 
\tau _H }{\tau _F - \tau _H}\right)^{\frac{4}{5}} .
\label{TempVol}
\een
where $R_C $ and $\tau_C$  are the final transverse and longitudinal   
sizes of the QGP; $v_C $ and  $a_C $ are its transverse flow velocity and 
transverse  acceleration at $\tau_C $; $T_C$ is the critical temperature of
the quark-hadron phase transition; $T_H $  is the temperature of the 
hadronic matter at the end of the mixed phase, occurring at the time 
$\tau_H $; and the kinetic freeze-out occurs at  $\tau _F $, when the    
temperature is $T_F $. We stress that the equation above is valid 
for $\tau \geq \tau _H $.  The above parametrization is an attempt to
mimic the hydrodynamic expansion of the  hadronic matter. In spite of its  
limitations it was useful in the study of the time evolution of the
multiplicity of other states (see more discussion in
Refs.\cite{ChoLee1,XProd2}).

%\item \textit{Fixing of the free parameters for central collisions.}

%%%%%%%%%%%%%%%%%%%%%%%%%%%%%%%%%%%%%%%%%%%%%%%%%%%%%%%%%%%%%
\subsubsection{Fixing the free parameters}
%%%%%%%%%%%%%%%%%%%%%%%%%%%%%%%%%%%%%%%%%%%%%%%%%%%%%%%%%%%%

We will address central $Pb-Pb$ collisions at $\sqrt{s_{NN}} = 5.02$ TeV 
at the LHC. Following Ref.~\cite{Abreu:2022lfy} the set of parameters in 
Eq. (\ref{TempVol}) is fixed as explained in ~\cite{ExHIC:2017smd} and is   
given in Table~\ref{param}. The total number of charm quarks in charmed
hadrons, $N_c$,  is conserved during the production and dissociation    
reactions.  This is enforced by the expression
$n_c(\tau) \times V(\tau) = N_c = const$, which leads to the time-dependent 
charm quark fugacity factor $\gamma _c $ in Eq.~(\ref{densities}). In the
case of light mesons, we use their fugacities as normalization parameters
to fit the multiplicities given in  Table~\ref{param}. 

%\begin{widetext}   
\begin{center}
\begin{table}[h!]
\caption{Parameters used in Eq.~(\ref{TempVol}) for the hydrodynamic
expansion of the hadronic medium formed in central $Pb-Pb$ collisions at
$\sqrt{s_{NN}} = 5.02$ TeV, and in Eq.~(\ref{coalmod}) for the coalescence
model~\cite{Abreu:2020ony,Abreu:2022lfy,ExHIC:2017smd}.}
\vskip1.5mm
\label{param}
\begin{tabular}{ c c c }
\hline
\hline
 $v_C$ (c) & $a_C$ (c$^2$/fm) & $R_C$ (fm)   \\   
%\hline
0.5 & 0.09 & 11  
\\  
\hline
%\hline
 $\tau_C$ (fm/c) & $\tau_H$ (fm/c)  &  $\tau_F$ (fm/c)  \\   
%\hline
7.1  & 10.2 & 21.5
\\  
\hline
%\hline
  $T_C \ (\MeV)$  & $T_H \ (\MeV)$ & $T_F \ (\MeV)$ \\   
%\hline
 156 & 156 & 115   \\  
\hline
%\hline
 $N_{\pi}(\tau_H)$ & $N_{K}(\tau_H)$ & $N_{\eta}(\tau_H)$ \\   
%\hline
 713 & 134 & 53 \\  
\hline
 $N_c$  & $N_s (\tau_H)$  & $N_{u}(\tau_H) (= N_{d}(\tau_H))$ \\   
%\hline
 14 & 386 & 700 \\  
\hline
 $m_c \ (\MeV)$ & $m_s \ (\MeV)$ & $m_q \ (\MeV)$ \\   
%\hline
1500  & 500 & 350 \\  
\hline
 $ \omega_c \ (\MeV)$ & & \\   
%\hline
220  &  &  \\  
% \hline
%%\hline
% \textcolor{red}{ $V_C $ (fm${}^3$) } & &  \\   
%%\hline
% \textcolor{red}{  5380}  &  &    \\  
\hline
\hline
\end{tabular}
\end{table}
\end{center}
%\end{widetext}

%\item \textit{Initial conditions via coalescence model.}

%%%%%%%%%%%%%%%%%%%%%%%%%%%%%%%%%%%%%%%%%%%%%%%%%%%%%%%%%%%%%%%%%
\subsubsection{Initial conditions via coalescence model}
%%%%%%%%%%%%%%%%%%%%%%%%%%%%%%%%%%%%%%%%%%%%%%%%%%%%%%%%%%%%%%%%% 

The yield of the $Z_{cs}$ state at the end of QGP is computed with the 
help of the coalescence model \cite{ExHIC:2017smd}.              
This approach is based on the overlap of the density matrix of its     
constituents with its Wigner function. It encodes some aspects of the
intrinsic structure of the system, such as angular momentum and the type 
and number of constituent quarks.  So, assuming that the $Z_{cs}$ state is 
a $S$-wave tetraquark with quark content $c \bar{c} s \bar{u} $, its
multiplicity at $\tau_C$ is given by
~\cite{ChoLee1,XProd2,Abreu:2020ony,Abreu:2022lfy,ExHIC:2017smd}:
\begin{eqnarray}
N_{Z_{cs}}^{coal} (\tau_C) & \approx & \frac{g_{Z_{cs}}
  [(4\pi)^3 M]^{3/2}}{ (\omega ^{3/2} V) ^3} \frac{ 1 }{ (1+ 2T/\omega)^{3}}
\, \,
\nonumber \\
& &\times  \frac{ N_c^2 N_s N_q }{g_c^2 g_s g_q (m_c^2 m_s m_q )^{3/2}}  
\label{coalmod}
\end{eqnarray}
where $g_j$ and $N_j$ are the degeneracy and number of the $j$-th     
constituent of the $Z_{cs}$ (the index $q$ refers to 
the light flavor quarks). The hadron is assumed to behave like a
harmonic oscillator and the quantity $\omega $ is the oscillator frequency.
The  frequency, the quark numbers and    
masses were taken from~\cite{Abreu:2022lfy,ExHIC:2017smd} and are
summarized in Table~\ref{param}. With the parameters from
Table~\ref{param}  Eq.~(\ref{coalmod}) yields
\begin{eqnarray}
N_{Z_{cs}}^{coal} (\tau_C) & \approx &   6.5 \times 10^{-7}. 
\label{coalmod2}
\end{eqnarray}
For the sake of comparison,  we compute the initial number of       
$Z_{cs}$ within the Statistical Hadronization Model (SHM), which is
based on Eq.~(\ref{densities}). We find  
$N_{Z_{cs}}^{stat} (\tau_H) \equiv  N_{Z_{cs}} (\tau_H) V(\tau_H)      
= 2.35 \times 10^{-3}$, which is about four orders of magnitude higher
than that obtained with the coalescence model. 

We emphasize that only the compact tetraquark configuration was explored.
In the molecular interpretation, the oscillation frequency is estimated 
to be $\omega = 6 B$, with $B$ being the binding energy. However, the 
observed $Z_{cs}$ mass is higher than the $ D_{s}^{* -} D^{0} $ or
$ D_{s}^{-} D^{* 0}  $ thresholds, which makes it difficult to
interpret the $Z_{cs}$ as a bound state of hadrons.

%%%%%%%%%%%%%%%%%%%%%%%%%%%%%%%%%%%%%%%%%%%%%%%%%%%%%%%%%%
\subsection{The system size dependence}
\label{size}
%%%%%%%%%%%%%%%%%%%%%%%%%%%%%%%%%%%%%%%%%%%%%%%%%%%%%%%%%%

In order to express the $Z_{cs}$ multiplicity in terms of measurable  
quantities, let us introduce  the dependence of the $N_{Z_{cs}}$ on 
the system size, represented here by the measurable central value of
the rapidity distribution of charged particles:
$\mathcal{N} = \left[ d N_{ch} / d \eta (|\eta| < 0.5)\right]^{1/3}$. 
To make predictions we will use empirical  relations which connect the
quantities listed in Table~\ref{param} with $\mathcal{N}$. The relations
are briefly described below. For more details, we refer the reader to
Ref.~\cite{Abreu:2022lfy}.

%%%%%%%%%%%%%%%%%%%%%%%%%%%%%%%%%%%%%%%%%%%%%%%%%%%%%%%%%%
\subsubsection{Kinetic freeze-out time}
%%%%%%%%%%%%%%%%%%%%%%%%%%%%%%%%%%%%%%%%%%%%%%%%%%%%%%%%%

The empirical formula relating $\mathcal{N}$ and the kinetic freeze-out
temperature $ T_F $ is given by~\cite{LeRoux:2021adw,Abreu:2022lfy}: 
\begin{equation}
T_F  = {T_{F0}} \, e^{- b \, \mathcal{N}},
\label{chiafit}
\end{equation}
where $T_{F0} = 132.5$ MeV and $ b = 0.02$. This parametrization has  
been chosen so as to fit the blastwave model analysis of the data    
performed by the ALICE Collaboration~\cite{alice13}. Inserting
Eq.~(\ref{chiafit}) into the Bjorken-like cooling relation
$ \tau_F T_F ^3 = \tau_H  T_H ^3 $, we obtain: 
\begin{equation}
\tau_F =  \tau_H \left( \frac{T_H}{T_{F0}} \right)^3 e^{3 b \mathcal{N}}.
\label{relf}
\end{equation}
Thus, the larger (smaller) the system and the multiplicity of produced 
hadrons, the longer (shorter) the duration of the hadron gas. For a given
system ($\mathcal{N}$), Eq.~(\ref{relf}) determines the time up to which we
integrate the rate equation~(\ref{rateeq}). Therefore $N_{Z_{cs}}$ will be 
a function of $\mathcal{N}$. 

%%%%%%%%%%%%%%%%%%%%%%%%%%%%%%%%%%%%%%%%%%%%%%%%%%%%%%%%%%%
\subsubsection{Volume}
%%%%%%%%%%%%%%%%%%%%%%%%%%%%%%%%%%%%%%%%%%%%%%%%%%%%%%%%%%% 

We start with the relation between the volume per rapidity
($dV/dy$) and $dN_{ch}/d\eta$ obtained in ~\cite{Vovchenko:2019kes}
with the SHM: 
\begin{equation}
  \frac{d V}{d y} = 2.4 \, \left.
  \frac{d N_{ch}}{d \eta}\right|_{ |\eta|<0.5}
= 2.4 \, \mathcal{N}^3.
\label{relVN0}
\end{equation}
The integration over the rapidity yields the relation                
$V \varpropto \mathcal{N}^3$, with $V$ being the chemical freeze-out  
volume $V_C  =  V (\tau_C) $. As in Ref.~\cite{ExHIC:2017smd}, we
assume that $ V_C = V_H = V (\tau_H)$.  The proportionality constant 
can be determined using the parametrization reported in  
Ref.~\cite{Niemi:2015voa} for $0-5\%$ centrality class in 5.02 TeV
$Pb-Pb$ collisions. It gives the volume
$ V_H = 5380 $ fm${}^3$ for
$\left[ d N_{ch} / d \eta (\eta < 0.5)\right]
= 1908$ ($\mathcal{N} \approx 12.43 $). 
Consequently, we obtain: 
\begin{equation}
V = 2.82 \, \mathcal{N}^3 .
\label{relV}
\end{equation}

%%%%%%%%%%%%%%%%%%%%%%%%%%%%%%%%%%%%%%%%%%%%%%%%%%%%%%%%%%%%%%%%
\subsubsection{Charm quark number}
%%%%%%%%%%%%%%%%%%%%%%%%%%%%%%%%%%%%%%%%%%%%%%%%%%%%%%%%%%%%%%%%

As we are not aware of any experimentally established connection between 
$N_c$ and $d N_{ch} / d \eta (\eta < 0.5)$ in the context of heavy ion
collisions, we make use of the data from the ALICE collaboration reported
in Ref.~\cite{ALICE:2015ikl} on the production of charm mesons in high 
multiplicity $pp$ collisions at $\sqrt{s} = 7$ TeV. The differential 
distribution of $D$ mesons as a function of $d N_{ch} / d \eta $  in
Fig. 2 of the mentioned paper may be parametrized by a power law,
which after the integration over the appropriate interval of 
rapidity and transverse momentum yields the relation
\cite{Abreu:2022lfy}:
\be
N_D  \varpropto  \left(\frac{d N_{ch}}{d \eta } \right)^{1.6}
     \varpropto \left(\mathcal{N}^3\right)^{1.6} , 
\label{relND}
\ee
We also assume that the charm quark number and the number of $ D $ mesons
are proportional: 
\be
N_c \varpropto N_D \varpropto \left(\mathcal{N}^3\right)^{1.6} . 
\label{propcD}
\ee
The proportionality constant can then be fixed by using the number shown in
Table~\ref{param} ($N_c = 14$ for $\mathcal{N} = 12.43$), yielding
\begin{equation}
N_c =  7.9 \times 10^{-5} \, \mathcal{N}^{4.8}.
\label{relNc}
\end{equation}

%\textcolor{red}{
%%%%%%%%%%%%%%%%%%%%%%%%%%%%%%%%%%%%%%%%%%%%%%%%%%%%%%%%%%%
\subsubsection{Light and strange quark numbers}
%%%%%%%%%%%%%%%%%%%%%%%%%%%%%%%%%%%%%%%%%%%%%%%%%%%%%%%%%%%
%\textcolor{red}{

In the case of the light and strange quark numbers, we follow the same 
procedure described in the previous subsection, i.e.,  making use of a
relation similar to  Eq.~(\ref{propcD}). In the lack of data relating 
directly $N_u, N_s$ to the charged particle multiplicity,  we use this 
power law with the exponent 1 rather than 1.6, taking into account the  
dependence of the pions and kaons with $d N_{ch} / d \eta $ reported in
Table ~3 of ~\cite{alice13}. Next, we can fix the proportionality constants
in order to match the numbers $N_u, N_s$ displayed in Table ~\ref{param}
at $\mathcal{N} = 12.43$, obtaining the expressions
\begin{eqnarray}
N_u & = &   0.37\, \mathcal{N}^{3}, \nonumber \\
N_s & = &  0.20  \, \mathcal{N}^{3}. 
\label{relNuNs}
\end{eqnarray}
%}

%%%%%%%%%%%%%%%%%%%%%%%%%%%%%%%%%%%%%%%%%%%%%%%%%%%%%%%%%%%%%%%
\subsubsection{Centrality and energy dependence}
%%%%%%%%%%%%%%%%%%%%%%%%%%%%%%%%%%%%%%%%%%%%%%%%%%%%%%%%%%%%%%

The quantity $\mathcal{N}$ depends on the ion mass number ($A$) on the
center-of-mass collision energy ($\sqrt{s}$) and on the centrality of
the collision. It is interesting to find the dependence of $\mathcal{N}$
on one of these variables, keeping the others constant. This was done in
Ref.~\cite{Niemi:2015voa}, where, fixing the energy at $5.02$ TeV and
choosing the projetiles to be $Pb-Pb$, the authors found (and displayed
in their Fig. 4) a relation between $\mathcal{N}$ and the centrality which
can be parametrized as:
\begin{eqnarray}
  \left. \frac{d N_{ch}}{d \eta}\right|_{ |\eta|<0.5} &  = & 2142.16 - 85.76
  x + 1.89 x^{2} - 0.03 x^{3} \nonumber \\ 
& &+3.67 \times 10^{-5} \ x^{4} - 2.24 \times 10^{-6} \ x^{5}   \nonumber \\ 
& &+ 5.25 \times 10^{-9} \ x^{6}, 
\label{relCN}
\end{eqnarray}
where $x$ denotes the centrality (in $\%$). Similarly, in the same paper 
we can extract the dependence of $\mathcal{N}$ on $\sqrt{s}$ from Fig. 3.  
It can be parametrized as: 
\begin{equation}
  \frac{d N_{ch}}{d \eta}    = -2332.12 + 491.69 \,
  \log ( 220.06 + \sqrt{s} )  
\label{relSN}
\end{equation}

%%%%%%%%%%%%%%%%%%%%%%%%%%%%%%%%%%%%%%%%%%%%%%%%%%%%%%%%%%%

\section{Results}
\label{Results}
%%%%%%%%%%%%%%%%%%%%%%%%%%%%%%%%%%%%%%%%%%%%%%%%%%%%%%%%%%%

%%%%%%%%%%%%%%%%%%%%%%%%%%%%%%%%%%%%%%%%%%%%%%%%%%%%%%%%%%%
\subsection{The time evolution}
\label{resultstime}
%%%%%%%%%%%%%%%%%%%%%%%%%%%%%%%%%%%%%%%%%%%%%%%%%%%%%%%%%%%

Here we present our results for the time evolution of the $Z_{cs}$
multiplicity by solving Eq.~(\ref{rateeq}). We emphasize that in our
calculation the $Z_{cs}$ is treated as a compact tetraquark. This is
consistent with the use of QCDSR (which are not appropriate to study 
extended hadron molecules) and with the choice of initial conditions 
given by the coalescence model with the parameters given in
Table~\ref{param}.
For the sake of comparison we will also show the results obtained with
the statistical hadronization model, i.e. with Eq.~(\ref{densities}). 

%%%%%%%%%%%%%%%%%%%%%%%%%%%%%%%%%%%%%%%%%%%%%%%%%%%%%%%%%%%%%%%%%%%%
%%%%%%%%%%%%%%%%%%%%%%%%%%%%%%%%%%%%%%%%%%%%%%%%%%%%%%%%%%%%%%%%%%%%
\begin{figure}[!ht]
\includegraphics[{width=1.0\linewidth}]{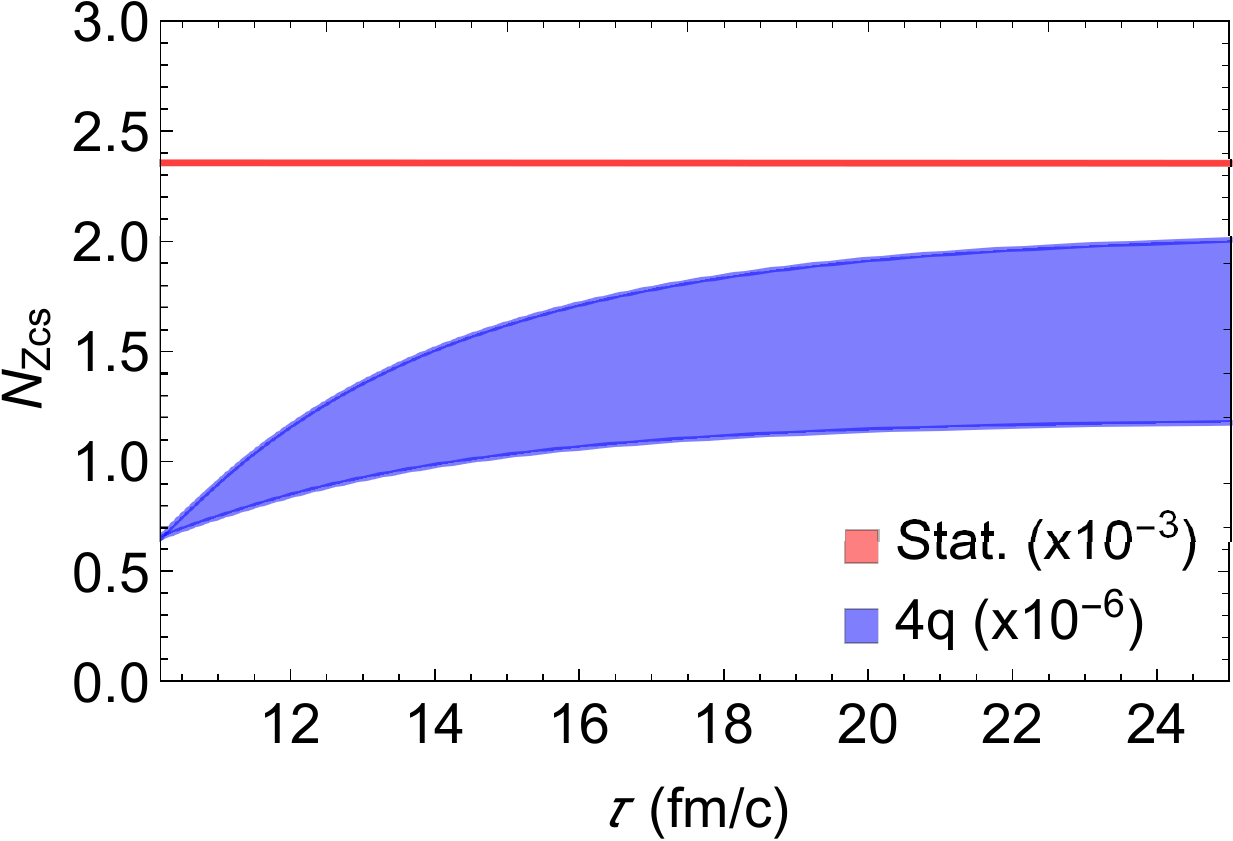}\\
\caption{$Z_{cs}$ multiplicity as a function of the proper time in
central $Pb-Pb$ collisions at $\sqrt{s_{NN}} = 5.02$ TeV.               
The curves represent the results obtained with initial conditions given 
by the coalescence and statistical hadronization models.
The band denotes the       
uncertainties coming from the QCDSR calculations of the absorption and
production cross section. 
%\textcolor{red}{Bottom panel: inclusion of the curve obtained with initial
%conditions given by the statistical model.} 
}
\label{TimeEvolNZcs}
\end{figure}
%%%%%%%%%%%%%%%%%%%%%%%%%%%%%%%%%%%%%%%%%%%%%%%%%%%%%%%%%%%%%%%%%%%%    
%%%%%%%%%%%%%%%%%%%%%%%%%%%%%%%%%%%%%%%%%%%%%%%%%%%%%%%%%%%%%%%%%%%%

In Fig.~\ref{TimeEvolNZcs} we show the  evolution of the $Z_{cs}$
abundance as a function of the proper time. The band represents the    
uncertainties coming from the QCDSR calculations of the absorption and
production cross sections. These results (obtained with initial conditions
given by the coalescence model) suggest that $ N_{Z_{cs}}$ increases by
a factor of $\simeq 2-3$ during the hadron gas phase. This behavior  is
the result of the competition between the two contributions on the right
side of the kinetic  equation~(\ref{rateeq}):  the gain and loss terms 
related to the  $Z_{cs}$-production and absorption  reactions,
respectively. The thermal  cross sections for  $Z_{cs}$ absorption are
bigger than those for  $Z_{cs}$ production. However, when multiplied by
$N_{Z_{cs}}$  (which is initially small) they become a small number  
and therefore the gain  term dominates, yielding a positive value for
the time derivative of $N_{Z_{cs}}^{coal}$. Hence,  the  $Z_{cs}$
multiplicity grows during the expansion and cooling of the system. 
The curve obtained with initial conditions given by  the Statistical 
Hadronization Model  remains practically constant during the hadronic gas
phase. This happens because the initial value $N_{Z_{cs}}^{stat}$ is much
higher than $N_{Z_{cs}}^{coal}$ and hence the loss term has the same
magnitude as the gain term. It is interesting to observe
that the approximate constancy of $N_{Z_{cs}}^{stat}$ is the practical
definition of chemical equilibrium.  In the SHM it is an assumption. Our
numerical calculations give some support to it.

%%%%%%%%%%%%%%%%%%%%%%%%%%%%%%%%%%%%%%%%%%%%%%%%%%%%%%%%
\subsection{$\mathcal{N}$, $\sqrt{s}$ and centrality dependence}
\label{resultssize}
%%%%%%%%%%%%%%%%%%%%%%%%%%%%%%%%%%%%%%%%%%%%%%%%%%%%%%%% 

Now we present and discuss our results for the $Z_{cs}$ multiplicity
as a function of  $\mathcal{N}$, of $\sqrt{s}$ and of the centrality. 
First we determine the $\mathcal{N}$-dependent initial conditions via      
coalescence model by substituting the relations (\ref{relV}), (\ref{relNc})  
and (\ref{relNuNs}) into (\ref{coalmod}).
%\textcolor{red}{
These substitutions yield the relation
\be
N_{Z_{cs}}^{coal} \propto \mathcal{N}^{6.6} .
\label{ndep}
\ee
%At this point we note a crucial difference between this result and the 
%$\mathcal{N}$  dependence obtained in Ref.~\cite{Abreu:2022lfy}. Here     
%we are considering the relations in (\ref{relNuNs}) between $\mathcal{N}$
%and light and strange quark numbers. Since these formulas were not included
%in our previous work,  the power law of the $\mathcal{N}$ 
%dependence for tetraquark configuration was found to be 
%$N^{4q} \propto \mathcal{N}^{0.6}$, rather than that of Eq.~(\ref{ndep}).
%}
Then, with the $\mathcal{N}$ dependent initial conditions, the kinetic    
equation (\ref{rateeq}) was integrated up to the kinetic freeze-out time,
$\tau_F$, which, in its turn,  also depends on  $\mathcal{N}$ according to
Eq.~(\ref{relf}).

%%%%%%%%%%%%%%%%%%%%%%%%%%%%%%%%%%%%%%%%%%%%%%%%%%%%%%%%%%%%%%%%%%%%
%%%%%%%%%%%%%%%%%%%%%%%%%%%%%%%%%%%%%%%%%%%%%%%%%%%%%%%%%%%%%%%%%%%%
\begin{figure}[!ht]
\includegraphics[{width=1.0\linewidth}]{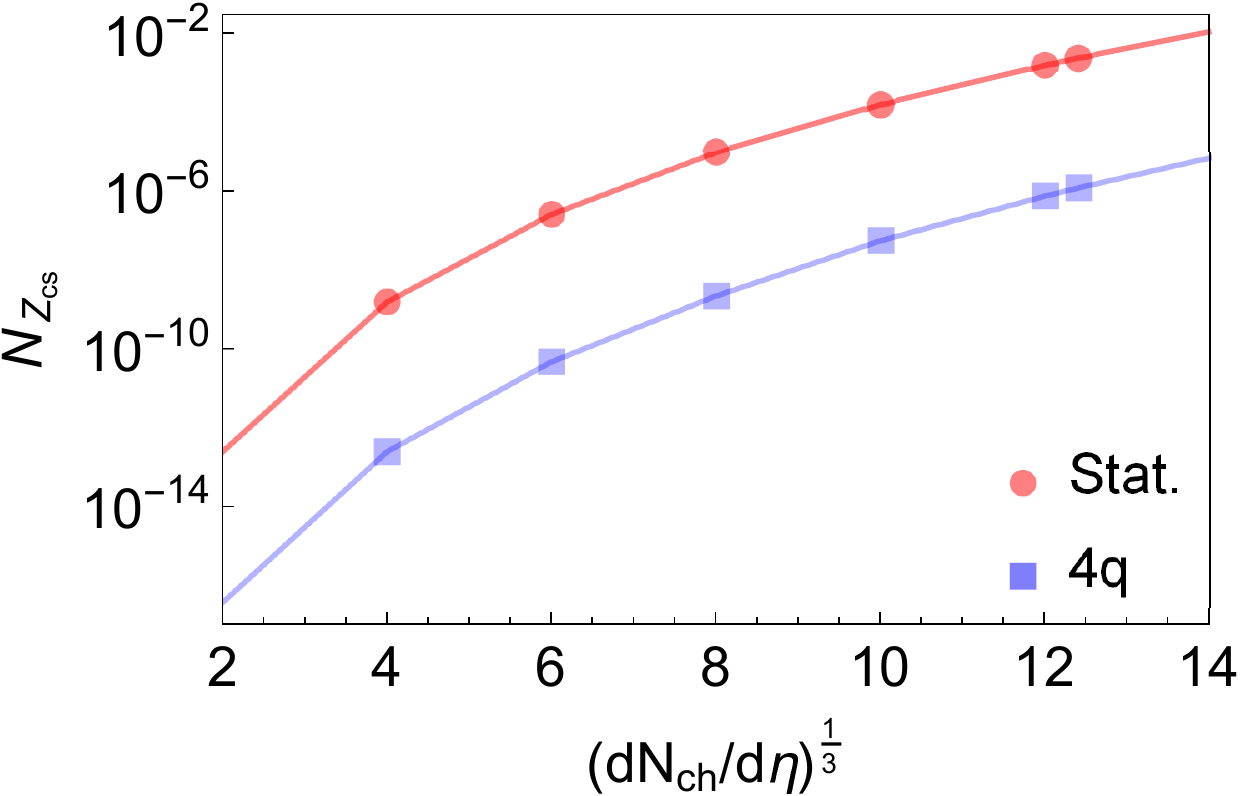}\\
\caption{The $Z_{cs}$ multiplicity as a function of $\mathcal{N}$. The  
curves represent the results obtained with initial conditions given by
the coalescence and statistical hadronization models.}
\label{Fig:NZcsdNdEta}
\end{figure}
%%%%%%%%%%%%%%%%%%%%%%%%%%%%%%%%%%%%%%%%%%%%%%%%%%%%%%%%%%%%%%%%%%%%    
%%%%%%%%%%%%%%%%%%%%%%%%%%%%%%%%%%%%%%%%%%%%%%%%%%%%%%%%%%%%%%%%%%%%

The method summarized above allows us to generate the plot of the $Z_{cs}$ 
abundance as a function of $\mathcal{N}$, shown in
Fig.~\ref{Fig:NZcsdNdEta}. The band representing the uncertainties is not 
visible here because the range of the abundances considered is much larger
than in the previous figure. 
We observe that the multiplicity increases as the system size grows.
Comparing the regions of p-p and Pb-Pb collisions                  
(for example $\mathcal{N} \sim  2-3$ and $10-12.5$, respectively),
$N_{Z_{cs}}$ changes several orders of magnitude. Both  
$N_{Z_{cs}}^{coal}$ and $N_{Z_{cs}}^{stat}$ show a similar behavior, 
but with different magnitudes. Thus, these results strongly suggest
that collisions involving heavy ions appear as a very  interesting
environment to investigate the  $Z_{cs}$ properties and discriminate
its intrinsic nature.

%%%%%%%%%%%%%%%%%%%%%%%%%%%%%%%%%%%%%%%%%%%%%%%%%%%%%%%%%%%%%%%%%%%%
%%%%%%%%%%%%%%%%%%%%%%%%%%%%%%%%%%%%%%%%%%%%%%%%%%%%%%%%%%%%%%%%%%%%
\begin{figure}[!ht]
\includegraphics[{width=1.0\linewidth}]{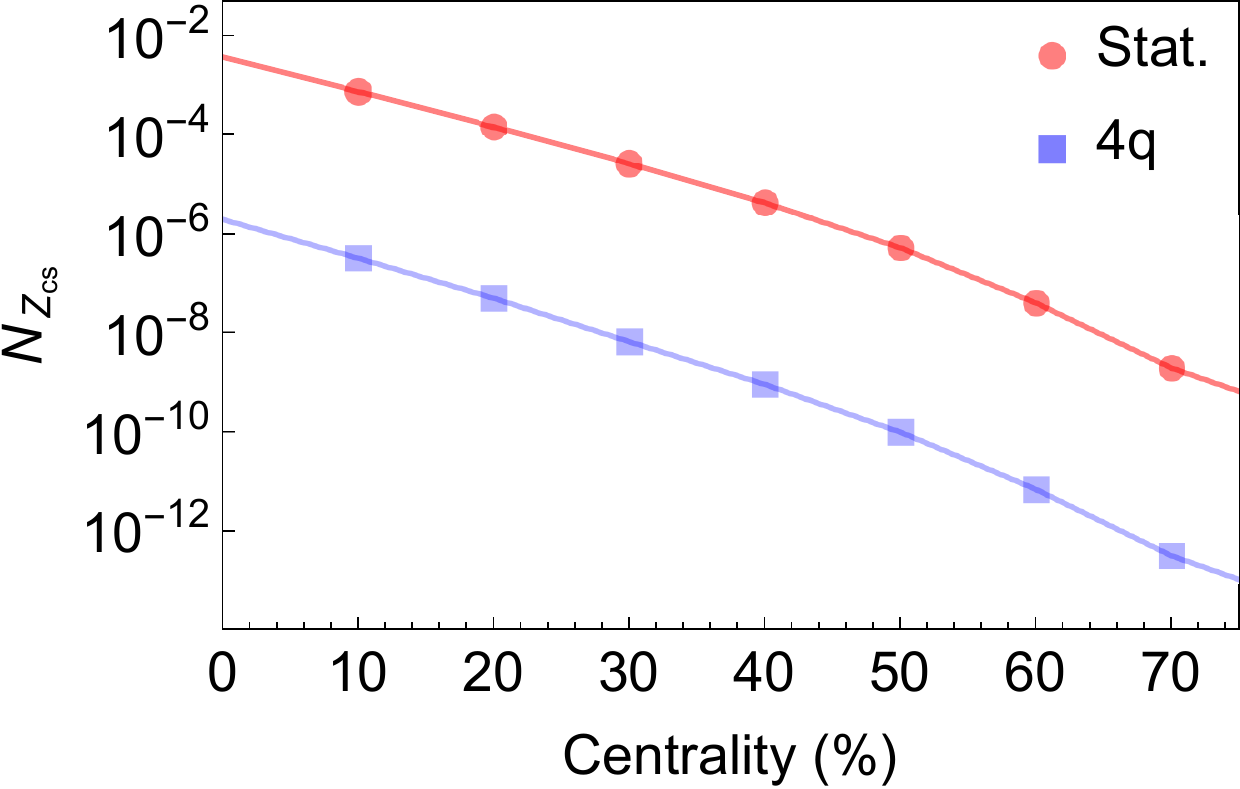}\\
\caption{The $Z_{cs}$ multiplicity as a function of centrality in 
$Pb-Pb$ collisions at $\sqrt{s_{NN}} = 5.02$ TeV. The curves represent
the results obtained with initial conditions given by the coalescence
and statistical hadronization models. }
\label{Fig:NZcsCentrality}
\end{figure}
%%%%%%%%%%%%%%%%%%%%%%%%%%%%%%%%%%%%%%%%%%%%%%%%%%%%%%%%%%%%%%%%%%%%    
%%%%%%%%%%%%%%%%%%%%%%%%%%%%%%%%%%%%%%%%%%%%%%%%%%%%%%%%%%%%%%%%%%%%

%%%%%%%%%%%%%%%%%%%%%%%%%%%%%%%%%%%%%%%%%%%%%%%%%%%%%%%%%%%%%%%%%%%%
%%%%%%%%%%%%%%%%%%%%%%%%%%%%%%%%%%%%%%%%%%%%%%%%%%%%%%%%%%%%%%%%%%%%
\begin{figure}[!ht]
\includegraphics[{width=1.0\linewidth}]{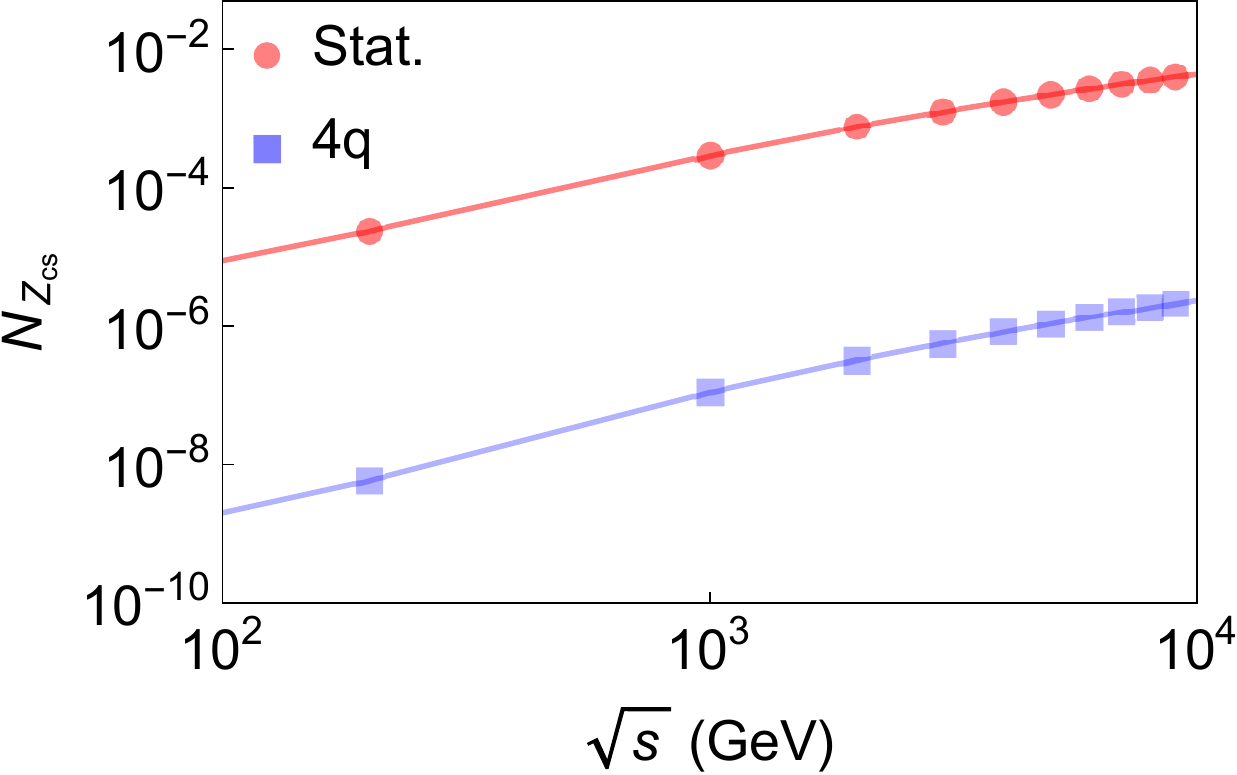}\\
\caption{The $Z_{cs}$ multiplicity as a function of the center-of-mass
  energy  $\sqrt{s_{NN}}$  in central $Pb-Pb$ collisions. The curves  
  represent the results obtained with initial conditions given by the
  coalescence and statistical hadronization models. }
\label{Fig:NZcsSqrts}
\end{figure}
%%%%%%%%%%%%%%%%%%%%%%%%%%%%%%%%%%%%%%%%%%%%%%%%%%%%%%%%%%%%%%%%%%%%    
%%%%%%%%%%%%%%%%%%%%%%%%%%%%%%%%%%%%%%%%%%%%%%%%%%%%%%%%%%%%%%%%%%%%

For completeness, we show the dependence of the $Z_{cs}$  
abundance with other relevant observables. Making use of the 
relations~(\ref{relCN}) and ~(\ref{relSN}), in Figs.~\ref{Fig:NZcsCentrality} 
and ~\ref{Fig:NZcsSqrts} we present  $N_{Z_{cs}}$  as a function of 
centrality and center-of-mass energy  $\sqrt{s_{NN}}$, respectively.    
They show the strong dependence of the multiplicity with the centrality
and $\sqrt{s_{NN}}$. The multiplicity $N_{Z_{cs}}$ increases with       
centrality by several orders of magnitude. In addition, our predictions 
for the $\sqrt{s_{NN}}$ dependence indicate that at LHC energies
(i.e. $1-10$ TeV) $N_{Z_{cs}}$
grows by  one order of magnitude. We expect that these estimates
can be tested experimentally in a near future.

%%%%%%%%%%%%%%%%%%%%%%%%%%%%%%%%%%%%%%%%%%%%%%%%%%%%%%%%%%%%%%%%%%%%
\section{Concluding remarks}
\label{Conclusions}
%%%%%%%%%%%%%%%%%%%%%%%%%%%%%%%%%%%%%%%%%%%%%%%%%%%%%%%%%%%%%%%%%%%%

In summary, we have improved our previous calculations \cite{Abreu:2022jmi} 
of the $Z_{cs}$ cross sections, introducing QCDSR form factors and
coupling constants. We have further continued the calculations, solving
the rate equation and determining the time evolution of the  $Z_{cs}$
multiplicity. The main conclusions is that $N_{Z_{cs}}$ grows by a factor
two during the hadron gas phase of heavy ion collisions at the LHC. Then,
using a series of empirical relations connecting several variables to the
measured central rapidity density, $\mathcal{N}$, we have made predictions
for the behavior of  $N_{Z_{cs}}$ with the system size, which can be
confronted with data, when they are available.

This work is part of a comprehensive effort to study the behavior of all
the new multiquark states in a hadron gas. This study is of crucial
importance, since in the near future these states
will be investigated  in relativistic heavy ion collisions.
Progress has been achieved and today we certainly know more than when this
program started,
ten years ago. But there is much more to be done. From now on, one of the
priorities will be to look for a more comprehensive approach, treating the
states as members of multiplets which share commmon properties, instead of
studying them one by one. In this sense, the classification shown in
Fig.~\ref{nonet} is welcome.

%%%%%%%%%%%%%%%%%%%%%%%%%%%%%%%%%%%%%%%%%%%%%%%%%%%%%%%%%%%%%%%%%%%%%%%%%%
\begin{acknowledgements}
%%%%%%%%%%%%%%%%%%%%%%%%%%%%%%%%%%%%%%%%%%%%%%%%%%%%%%%%%%%%%%%%%%%%%%%%%%

  The authors would like to thank the Brazilian funding agencies for their
  financial support: CNPq,
  FAPESB and INCT-FNA.  

%%%%%%%%%%%%%%%%%%%%%%%%%%%%%%%%%%%%%%%%%%%%%%%%%%%%%%%%%%%%%%%%%%%%%%%%%%%

\end{acknowledgements} 
%%%%%%%%%%%%%%%%%%%%%%%%%%%%%%%%%%%%%%%%%%%%%%%%%%%%%%%%%%%%%%%%%%%%%%%%%%%

\end{document}